\documentclass[11pt]{amsart}
\ifx\undefined\cal\let\cal=\mathcal\fi
\usepackage{amsmath}
\usepackage{amssymb}
\usepackage{graphicx}
\usepackage{hyperref}
\newcommand{\bfm}[1]{\mbox{\boldmath ${#1}$}}
\newcommand{\nonum}{\nonumber \\}
\newcommand{\beqa}{\begin{eqnarray}}
	\newcommand{\eeqa}[1]{\label{#1}\end{eqnarray}}
\newcommand{\beq}{\begin{equation}}
	\newcommand{\eeq}[1]{\label{#1}\end{equation}}
\newcommand\eq[1] {(\ref{#1})} 
\newcommand{\Grad}{\nabla}
\newcommand{\Div}{\nabla \cdot}
\newcommand{\Curl}{\nabla \times}

\newcommand{\Real}{\mathop{\rm Re}\nolimits}
\newcommand{\Imag}{\mathop{\rm Im}\nolimits}
\newcommand{\Tr}{\mathop{\rm Tr}\nolimits}

\newcommand{\lang}{\langle}
\newcommand{\rang}{\rangle}
\newcommand{\Md}{\partial}

\newcommand{\dund}[1]{\underline{\underline{#1}}}

\newcommand{\Ga}{\alpha}
\newcommand{\Gb}{\beta}
\newcommand{\Gd}{\delta}

\newcommand{\Gf}{\phi}

\newcommand{\Gg}{\gamma}
\newcommand{\Gc}{\chi}

\newcommand{\Gk}{\kappa}

\newcommand{\Gl}{\lambda}

\newcommand{\Gm}{\mu}

\newcommand{\Gt}{\theta}

\newcommand{\Gs}{\sigma}

\newcommand{\Go}{\omega}

\newcommand{\Gy}{\psi}

\newcommand{\BGe}{\bfm\epsilon}

\newcommand{\BGc}{\bfm\chi}

\newcommand{\BGs}{\bfm\sigma}

\newcommand{\BGG}{\bfm\Gamma}
\newcommand{\BGL}{\bfm\Lambda}

\newcommand{\BGY}{\bfm\Psi}

\newcommand{\CA}{{\cal A}}

\newcommand{\CD}{{\cal D}}
\newcommand{\CE}{{\cal E}}

\newcommand{\CJ}{{\cal J}}

\newcommand{\BCE}{{\bfm{\cal E}}}

\newcommand{\BCH}{{\bfm{\cal H}}}

\newcommand{\BCJ}{{\bfm{\cal J}}}

\newcommand{\BCR}{{\bfm{\cal R}}}
\newcommand{\BCS}{{\bfm{\cal S}}}

\newcommand{\BCU}{{\bfm{\cal U}}}
\newcommand{\BCV}{{\bfm{\cal V}}}
\newcommand{\BCW}{{\bfm{\cal W}}}

\newcommand{\bpm}{\begin{pmatrix}}
\newcommand{\epm}{\end{pmatrix}}

\def\Ba{{\bf a}}

\def\Be{{\bf e}}

\def\Bh{{\bf h}}

\def\Bj{{\bf j}}
\def\Bk{{\bf k}}

\def\Bn{{\bf n}}

\def\Bq{{\bf q}}
\def\Br{{\bf r}}

\def\Bt{{\bf t}}
\def\Bu{{\bf u}}
\def\Bv{{\bf v}}
\def\Bw{{\bf w}}
\def\Bx{{\bf x}}

\def\Bz{{\bf z}}
\def\BA{{\bf A}}
\def\BB{{\bf B}}
\def\BC{{\bf C}}

\def\BE{{\bf E}}

\def\BI{{\bf I}}
\def\BJ{{\bf J}}
\def\BK{{\bf K}}
\def\BL{{\bf L}}

\def\BP{{\bf P}}
\def\BQ{{\bf Q}}
\def\BR{{\bf R}}
\def\BS{{\bf S}}
\def\BT{{\bf T}}
\def\BU{{\bf U}}

\def\half{{\scriptstyle{1\over 2}}}

\numberwithin{equation}{section}

\title[Approximating the elasticity tensor of 2-d polycrystals as a function of the crystal moduli ]{A continued fraction approximation for the effective elasticity tensor of two-dimensional polycrystals as a function of the crystal elasticity tensor}

\author[G.W. Milton]{Graeme W. Milton}

\address[G.W.Milton]{Department of Mathematics, The University of Utah, Utah, USA}
\email{{\tt graeme.milton@utah.edu}}

\keywords{Homogenization, Elasticity, Continued Fraction Approximations, Two-Dimensional Polycrystals, Effective Tensors}

\subjclass[2020]{74Q05}

\begin{document}

%\title{A continued fraction approximation for the effective elasticity tensor of two-dimensional polycrystals as a function of the crystal elasticity tensor}
%\author{Graeme W. Milton}
%\date{September 10th, 2023}
%\maketitle
\begin{abstract}
For two-dimensional polycrystals the effective elasticity tensor $\BC_*$ as a function $\BC_*(\BC_0)$ of the elasticity tensor $\BC_0$ of the constituent crystal is considered.
It is shown that this function can be approximated by one with a continued fraction expansion resembling that associated with a class of microstructure known as sequential laminates. These are hierarchical microstructures defined inductively. Rank 0 sequential laminates are simply rotations of the pure crystal. Rank $j$ sequential laminates are obtained by laminating together,  on a length scale much larger that the existing microstructure and with interfaces perpendicular to some direction $\Bn_j$, rank $j-1$ sequential laminates with a rotation of the pure crystal. The continued fraction 
approximation for arbitrary polycrystal microstructures typically takes a more general form than that of sequential laminates, but has some free parameters. It is an open question as to whether these free parameters can always be adjusted so the continued fraction 
approximation matches exactly that of  a sequential laminate. If so, one would have established that the elastic response of two-dimensional polycrystals can always be mimicked by that of sequential laminates. Our analysis carries over to the 
more general case
where the strain is replaced by a field $\BE(\Bx)$ that is the gradient of a vector potential 
$\Bu(\Bx)$, i.e.  $\BE=\Grad\Bu$ and the stress is replaced by a 
matrix valued field $\BJ(\Bx)$ that need not be symmetric but has zero divergence 
$\Div\BJ=0$. The tensor $\BL(\Bx)$ entering the constitutive relation $\BJ=\BL\BE$
is locally a rotation of the tensor $\BL_0$ of the pure crystal that need not have any special symmetries and has 16 independent tensor elements. 
\end{abstract}	

\maketitle

%%%%%%%%%%%%%%%%%%%%%%%%%%%%%%%%%%%%%%%%%%%%%%%%%%%%%%%%%%%%%%%%%%%%%%%%
\section{Introduction}
%%%%%%%%%%%%%%%%%%%%%%%%%%%%%%%%%%%%%%%%%%%%%%%%%%%%%%%%%%%%%%%%%%%%%%%%%%%%%
\setcounter{equation}{0}

It is a pleasure writing this article in honor of Luc Tartar who, beside many other works, made pioneering contributions to the subject of homogenization theory, to bounds on the effective tensors that according to homogenization theory  govern the macroscopic response of composites, and to obtaining formulas for the effective tensors of laminates, often in collaboration with Francois Murat. 

This article concerns two-dimensional polycrystals, a composite of grains of a single anisotropic pure crystal rotated according to the grain orientation, as illustrated in Figure 1. Our primary interest is the function $\BC_*(\BC^{(0)})$ giving the effective elasticity tensor $\BC_*$ as a function of the elasticity tensor $\BC^{(0)}$ of the pure crystal. The goal is to show that it can be approximated to an arbitrarily high degree of approximation by a continued  fraction having a close relation with the continued fraction for microstructures known as sequential laminates, which will be described in the next section. After approximating the Hilbert space to a finite dimensional one,
the crux of our analysis will be a dimension counting argument showing that certain subspaces have a non-trivial intersection.  This reduces the problem to a simpler one, where the dimension counting argument can be employed again, and so forth, until one arrives at an elementary problem that is easily solved. The effective tensors at successive stages are linked, and elimination of the intermediate effective tensors gives rise to the continued fraction.

To be clear, it still remains an open question as to whether in general  $\BC_*(\BC^{(0)})$ can always be approximated by the function for some sequential laminate, rather than only by continued  fractions having a more general form.
Our continued fraction expansion provides a representation of   $\BC_*(\BC^{(0)})$ . Besides the intrinsic interest, this itself is useful as representation formulas
for effective tensors as a function of the tensors of the constituent materials allow one to obtain bounds on the response of
composites: see, for example, Chapters 18, 27, 28 in \cite{Milton:2002:TOC} and 
accompanying references. There is a connection between many of these bounds and bounds on Herglotz and Stieltjes functions. These have a long history  dating back to works of Nevanlinna and Pick: see \cite{Krein:1977:MMP} and references therein. 

\begin{figure}[!ht]
	\includegraphics[width=0.9\textwidth]{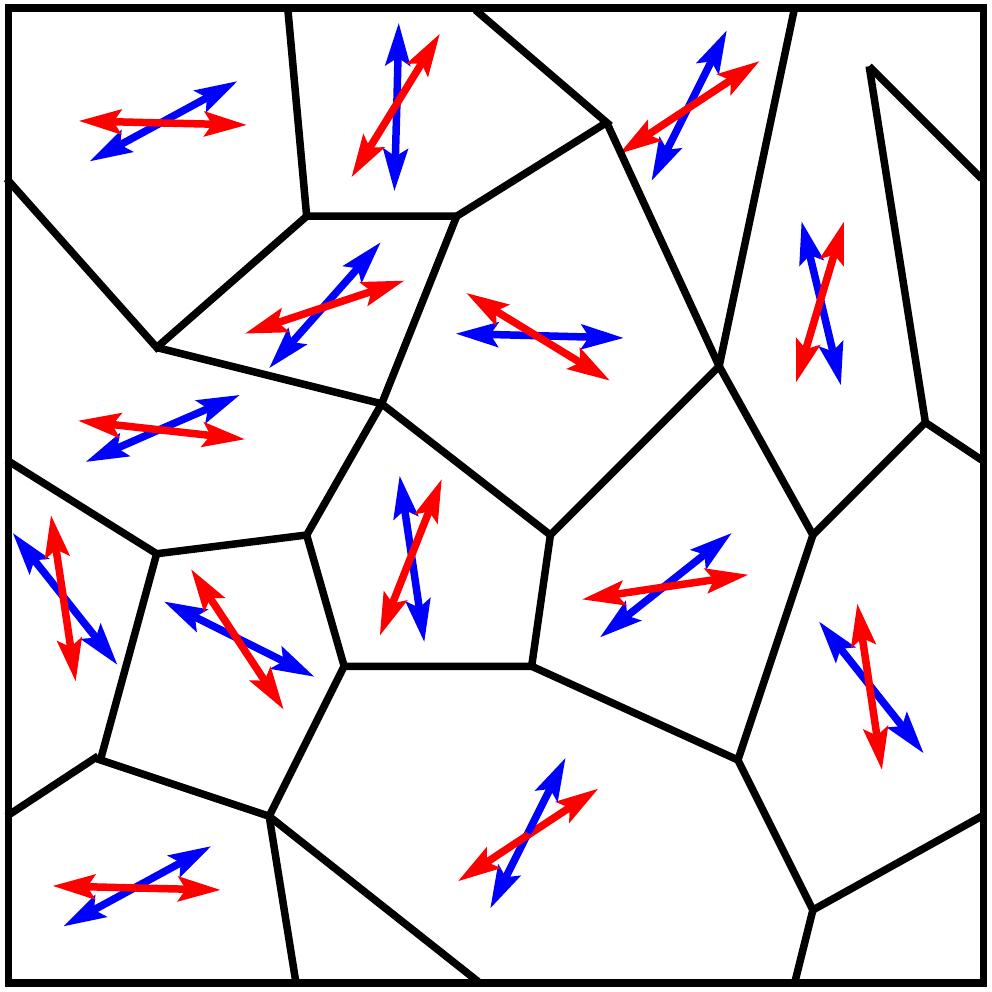}
	\caption{An example of a periodic two dimensional polycrystal showing the unit cell $Q$ of periodicity.
		 The double arrowed blue lines denote the 
     the direction of the non-trivial local eigenvector of the rank-one matrix valued function $\chi$ relative to which
		the crystal orientation is determined. We are free to rotate these eigenvectors by a common angle 
		while keeping $\BL(\Bx)$ unchanged: the coefficients $L^{(0)}_{ij}$ need to be adjusted accordingly.}
	\end{figure}

Without loss of generality one may assume the rotation fields, and hence
the fourth order elasticity tensor $\BC(\Bx)$ are periodic functions of $\Bx$ with unit cell $Q$. It takes the form
\beq \BC(\Bx)= \BR(\Bx)\BR(\Bx)\BC^{(0)}\BR^T(\Bx)\BR^T(\Bx),
\eeq{b.1}
in which $\BC^{(0)}$ is the elasticity tensor of the pure crystal, and $\BR(\Bx)$ is a $Q$-periodic $2\times 2$ matrix valued rotation field satisfying
\beq \BR(\Bx)[\BR(\Bx)]^T=\BI. \eeq{b.1a}
Note that here we are allowing for polycrystals more general that those in Figure 1: the rotation field  $\BR(\Bx)$ need not be piecewise constant but, for example, could vary continuously with $\Bx$.
The underlying equations are
\beq \Div\BGs=0,\quad \BGs(\Bx)=\BC(\Bx)\BGe(\Bx),\quad\BGe=[\Grad\Bu+(\Grad\Bu)^T)]/2, \eeq{b.2}
where $\BGs$ is the stress, $\BGe$ is the strain, both $Q$-periodic, and $\Bu$ is the displacement field, that may have an affine component in addition to a $Q$-periodic component. Given any prescribed value of
the applied strain $\langle\BGe\rangle$, where the angular brackets $\lang\cdot\rang$ denote a volume average over the cell of periodicity, the equations \eq{b.2} have a unique solution for $\BGs$, $\BGe$ and up to an additive constant for $\Bu$
provided $\BC^{(0)}$ is positive definite on the space of symmetric matrices (or provided $\Imag(\BC^{(0)})$ is positive definite if one is considering viscoelasticity in the quasistatic limit where $\BC_*$ may take complex values). The average stress
$\lang\BGs\rang$ depends linearly on the applied average strain $\lang\BGe\rang$ and it is this linear relation that defines the effective elasticity  tensor $\BC_*$:%
\beq  \lang\BGs\rang=\BC_*\lang\BGe\rang. \eeq{b.3}
According to homogenization theory $\BC_*$ determines the macroscopic response of the polycrystal to slowly varying applied 
fields where the length scale of variation is much larger than the periodic microstructure. We will consider a more general  problem where \eq{b.1}, \eq{b.2}, and \eq{b.3} are replaced by the equations
\beqa \BL(\Bx)& = &   \BCR(\Bx)\BR(\Bx)\BL^{(0)}\BR^T(\Bx)\BCR^T(\Bx),
\nonum
 \Div\BJ& = & 0,\quad \BJ(\Bx)=\BL(\Bx)\BE(\Bx) , \quad \BE=\Grad\Bu, 
\eeqa{b.4}
where $\BL^{(0)}$, in contrast to $\BC^{(0)}$ need not be self-adjoint nor have  antisymmetric matrices in its kernel.
Here we have chosen to distinguish  the rotation 
field $\BCR^T(\Bx)$ acting on the displacement field from
the rotation field $\BR(\Bx)^T$ acting  on the gradient and analogously for the rotation fields acting on $\BL^{(0)}\BR^T(\Bx)\BCR^T(\Bx)$.  These are still
rotations by the same angle $\Gt(\Bx)$  and satisfy
\beq \BR(\Bx)[\BR(\Bx)]^T=\BI ,\quad
 \BCR(\Bx)[\BCR(\Bx)]^T=\BI .
\eeq{b.4a}
Again these equations have a unique solution for the fields $\BJ(\Bx)$ and $\BE(\Bx)$ for any given
applied field $\lang\BE\rang$  when  the self-adjoint part of $\BC^{(0)}$ is positive definite. Then
$\lang\BJ\rang$ depends linearly on $\lang\BE\rang$ and it is this linear relation
\beq \lang\BJ\rang  =  \BL_*\lang\BE\rang  \eeq{b.4aaa}
that defines the effective tensor $\BL_*$. 
To reduce this to the elasticity problem one may simply set $\BL^{(0)}=\BC^{(0)}$.  Of course then $\BL^{(0)}$ is not positive definite. Nevertheless we can obtain an equivalent problem by using appropriate "translations" to shift $\BL^{(0)}$ to a positive definite tensor, with the effective tensor $\BL_*$ undergoing the same shift.  (The quadratic form associated with such a "translation" is known as a null-Lagrangian.) Specifically, we may define a fourth order tensor
$\BT$ whose action is to rotate a field $\BA(\Bx)$ by $90^\circ$:
\beq \BT\BA(\Bx)=\BR_{\perp}\BCR_{\perp}\BA(\Bx)=\BI\Tr[\BA(\Bx)]-\BA(\Bx)^T,
\eeq{b.7a}
where $\Tr$ denotes the trace of a matrix and
$\BR_{\perp}$ and $\BCR_{\perp}$ can be equated with the 
rotations $\BR(\Bx)$ and $\BCR(\Bx)$ when  $\Gt(\Bx)=\pi/2$. 
Then, since $\BT\BE$ is divergence free if $\BE$ is the gradient of a potential, it follows that if we shift $\BC^{(0)}$ by adding a  multiple $c$ of $\BT$ to it giving  $\BL^{(0)}=\BC^{(0)}+c\BT$, then $\BL_*$ will shift in the same way, giving  $\BL_*=\BC_*+c\BT$,  An appropriate choice of the multiple of $\BT$ leads to an equivalent problem with a shifted $\BL^{(0)}$ that is positive definite. The proof of this is a straightforward extension of, for example, the proof  in Section 6.4 of \cite{Milton:2002:TOC}.

The function  $\BL_*(\BL^{(0)})$ has the properties:
 \vskip4mm

$\bullet$ Analyticity: $\BL_*(\BL^{(0)}) $  is an analytic function of  $\BL^{(0)}$ on at least the union over  $\Gf\in [0,2\pi) $ of the domains  where the self-adjoint part of $\Real[e^{i\Gf}\BL^{(0)}]$ is
positive definite. 
 \vskip4mm

$\bullet$ Homogeneity: $\BL_*(\Gl\BL^{(0)})=  \Gl\BL_*(\BL^{(0)})$ for all complex $\Gl$.
 \vskip4mm

$\bullet$ Normalization: $\BL_*(\BI)=\BI$.
 \vskip4mm

$\bullet$ Herglotz Property: The self-adjoint part of $ \Imag[\BL_*(\BL^{(0)})]$   is positive definite when the  self-adjoint part of   $\Imag[\BL^{(0)}]$ is positive definite. 
 \vskip4mm

%\beqa &~&\text{Analyticity:} \quad \BL_*(\BL^{(0)}) \text{ is an analytic function of  } \BL^{(0)} \text{ on the %union over  } \Gf\in [0,2\pi) \nonum
%&~&\quad\quad\quad\text{ of the domains  where}
%\Imag[e^{i\Gf}\BL^{(0)}]>0. \nonum
%&~&\text{Homogeneity:}\quad \BL_*(\Gl\BL^{(0)})=  \Gl\BL_*(\BL^{(0)}),\quad\text{for all complex }\Gl,
%\nonum
%&~&\text{Normalization:}\quad  \BL_*(\BI)=\BI \nonum
%&~&\text{Herglotz Property:}\quad \Imag[\BL_*(\BL^{(0)})]>0 \text{       when    } \Imag[\BL^{(0)}]>0 .
%\eeqa{b.7}

Here we essentially provide a representation of these functions $\BL_*(\BL^{(0)})$. Namely, to an arbitrarily high degree of approximation, they can be approximated by
continued fractions having a similar form as that associated with sequential laminates. At first sight this seems like an enormously challenging problem since
$\BL_*(\BL^{(0)})$ is a function of 16 parameters, namely the components of the matrix representing $\BL^{(0)}$.
A reduction in the number of variables to 13 is possible, using "translations".  Specifically, we may 
shift  $\BL^{(0)}$ by adding to it a multiple of the translation $\BT$ defined by \eq{b.7a}, and then $\BL_*$ will shift in the same way. Conversely, since $\BT\BJ$ is the gradient of a potential if $\BJ$ is divergence free, then if we shift $[\BL^{(0)}]^{-1}$ by adding a multiple of $\BT$ then $\BL_*^{-1}$ will shift in the same way.  In
other words we have 2 parameters at our disposal such that the dependence of $\BL_*$ on these parameters is straightforward. Also, due to homogeneity, if we multiply $\BL^{(0)}$ by a constant, then  $\BL_*$ will be multiplied by that same constant. This leaves $\BL_*$ having a non-trivial dependence on 13 parameters. 

\section{Motivation} 
%%%%%%%%%%%%%%%%%%%%%%%%%%%%%%%%%%%%%%%%%%%%%%%%%%%%%%%%%%%%%%%%%%%%%%%%%%%%%
\setcounter{equation}{0}

Some background is needed to motivate this study. As composites can exhibit different properties and sometimes strikingly different properties than their constituent materials, there is considerable interest in establishing the range of properties a composite may have  and identifying microstructures that achieve desired properties in this range.
A beautiful example is that of  Murat and Tartar \cite{Murat:1985:CVH} and independently Lurie and Cherkaev \cite{Lurie:1986:EEC}, who completely characterized the set of effective conductivity tensors $\BGs_*$, allowing for anisotropic ones, of  three-dimensional composites of two isotropic conducting phases mixed in fixed proportions. They obtained bounds and then showed that any effective conductivity tensor compatible with the bounds is realized by a hierarchical laminate geometry whose conception goes back to Maxwell \cite{Maxwell:1954:TEM}.
Hierarchical laminates are multiscale structures defined recursively. At the first stage one defines the constituent phases as rank 0 laminates.  To obtain a hierarchical laminate of rank $r\geq 1$ one laminates together  laminates of rank $r-1$ with laminates of rank at most $r-1$.  The direction of lamination varies at the different stages of lamination, and there is a wide separation of length scales between laminations so one can use the effective tensors of the constituent hierarchical laminates
to obtain the effective tensor at the next stage. Alternatively, any effective conductivity tensor attaining the bounds of Tartar, Murat, Lurie and Cherkaev, is realized by an assemblage of coated ellipsoids, a geometry whose conception goes back to the papers \cite{Milton:1980:BCD, Milton:1981:BCP} and which generalizes the coated sphere assemblage construction of Hashin \cite{Hashin:1962:EMH} and Hashin and Shtrikman \cite{Hashin:1962:VAT}. Assemblages of doubly coated ellipsoids attain points inside the bounds.

A constant theme is that in many instances hierarchical laminates can have unusual properties and often achieve bounds. In fact, it is hard to identify microstructures whose properties of interest cannot be replicated by hierarchical laminates built from 
the same constituent materials. Personally, I initially thought it was impossible to find hierarchical laminates having a Poisson ratio arbitrarily close to $-1$ until I found one incorporating chevron type elements \cite{Milton:1992:CMP}
(a related construction was later independently discovered by Larsen, Sigmund, and Bouwstra \cite{Larsen:1997:DFC} and has been used as a protective layer in Bontrager bicycle helmets, where the negative Poisson ratio assists in getting the desired positive curvature of the layer). Similarly, I had thought \cite{Milton:2021:SOP} it impossible 
to find hierarchical laminates that achieve a sign change of the effective Hall coefficient, a counterexample to the standard belief that the sign of the charge carriers determines  the sign of the Hall coefficient. Certain interlocking ring structures were proven to have the sign inversion property \cite{Briane:2009:HTD}, then the design was simplified and verified numerically
\cite{Kadic:2015:HES} and experimentally \cite{Kern:2016:EES}. An alternative knotted structure having the sign inversion property was later conceived \cite{Kern:2018:THE}. Much to my surprise, Christian Kern found a hierarchical laminate exhibiting the sign change \cite{Kern:2023:SREtemp}. On the other hand, an example of Sverak  \cite{Sverak:1992:ROC} has led to
a seven phase composite whose effective elasticity tensor cannot be replicated by a hierarchical laminate composed from the same phases (Section 31.9 in \cite{Milton:2002:TOC}) and Grabovsky has an example of an exact relation for effective tensors that is satisfied by  hierarchical laminate geometries, but not for some microstructures \cite{Grabovsky:2017:MIF}.

A much deeper question is to look at the effective tensor as a function of the tensors, or relevant moduli, of the constituent phases and ask whether this effective tensor function for an arbitrary microstructure can always be replicated by  a suitable
hierarchical laminate microstructure. An example is the proof \cite{Milton:1986:APLG} that sequential laminates  can reproduce to an arbitrarily high degree of approximation the effective conductivity tensor $\BGs_*$ as a function $\BGs_*(\Gs_1,\Gs_2)$ of the conductivities $\Gs_1$ and $\Gs_2$ of two isotropic phases for two-dimensional, possibly anisotropic composites of  these phases.  Sequential laminates are a subclass of hierarchical laminates, again defined recursively: one begins with a trivial sequential laminate which is simple laminate of the constituent phases
then at each successive stage one laminates one of the constituent phases with a sequential laminate obtained at a previous stage.
Key to this correspondence was that the function  $\BGs_*(\Gs_1,\Gs_2)$  in two dimensions satisfies 
Keller's phase interchange identity\cite{Keller:1964:TCC}.
\beq \BGs_*(\Gs_2, \Gs_1)=\Gs_1\Gs_2\BGs_*(\Gs_1,\Gs_2)/\det\BGs_*(\Gs_1,\Gs_2), \eeq{0.1}
in addition to analyticity, homogenity, normalization, and Herglotz properties analogous to those mentioned in the last section. 
Related continued fraction expansions \cite{Milton:2023:TCP} can be obtained more generally without assuming \eq{0.1}, though in this case one cannot generally make a correspondence with sequential laminate geometries. 

There are also related results concerning two-dimensional conducting polycrystals, comprised of grains of anisotropic pure crystal with conductivity $\BGs^{(0)}$ rotated according to the grain orientation.  Dykhne \cite{Dykhne:1970:CTD} showed that the effective tensor $\BGs_*$ of the polycrystal satisfies $\det\BGs_*=\det\BGs^{(0)}$. In addition, if $\Gl^+$, $\Gl^-$ are the maximum and minimum eigenvalues of $\BGs^{(0)}$ so that $\Gl^+\BI\geq \BGs^{(0)}\geq \Gl^-\BI$ 
then elementary variational principles imply $\Gl^+\BI\geq \BGs_*\geq \Gl^-\BI$ . Putting these together,  if $\Gl^+_*$, $\Gl^-_*$ are the maximum and minimum eigenvalues of $\BGs_*$  then these satisfy $\Gl^+_*\Gl^-_*=\Gl^+\Gl^-$ and 
$\Gl^-_*\leq\Gl^+_*\leq\Gl^+$. This is fact characterizes all possible effective conductivity tensors $\BGs_*$. It is easy to see that a 
simple laminate of the crystal with conductivity $\BGs^{(0)}$ layered in the direction of the eigenvector of  $\BGs^{(0)}$ with the same crystal rotated by $90^\circ$ achieves any pair of  $\Gl^+_*$, $\Gl^-_*$ compatible with these constrains. (Subsequently a complete characterization was 
given of the possible effective conductivity tensors $\BGs_*$ of two-dimensional polycrystals obtained from $n$-phases with orientations that may vary from grain to grain: see Section 22.5 of \cite{Milton:2002:TOC}, references therein, and \cite{Francfort:2009:POB}). 
A deeper result was obtained by  Clark and Milton \cite{Clark:1994:MEC} and Clark \cite{Clark:1997:CFR}  who  showed 
the underlying Hilbert space structure, and hence the effective conductivity tensor $\BGs_*$ as a function $\BGs_*(\BGs^{(0)})$ of  the pure crystal conductivity tensor $\BGs^{(0)}$ could be mimicked to an arbitrary degree of approximation by sequential laminates, as for example illustrated in Figure 2. Sequential laminates are a subset of hierarchical laminates: at the first stage one still defines the constituent phases as rank 0 laminates, and then to obtain a sequential laminate of rank $r\geq 1$ one laminates sequential laminates of rank $r-1$ with one or more of the constituent 
phases, possibly with an orientation that depends on $r$.  
 This correspondence then led to a continued fraction expansion that approximates the function  $\BGs_*(\BGs^{(0)})$ .

\begin{figure}[!ht]
	\includegraphics[width=0.9\textwidth]{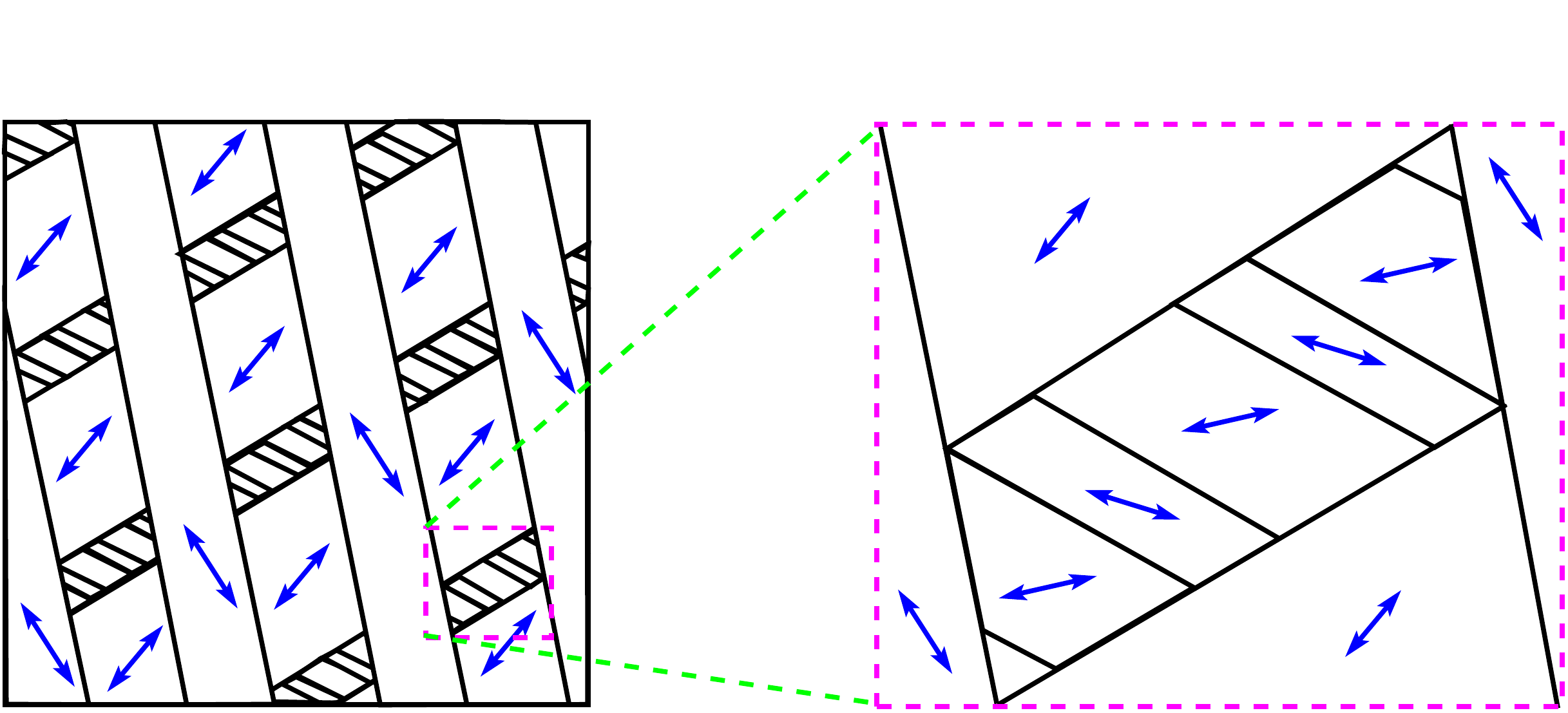}
	\caption{An example of a rank 3  two dimensional sequential laminate polycrystal.  It is schematic in the sense that there should be a large separation of length scales at each stage of the lamination. Here the crystal orientation is controlled by the function $\Gc$ which takes rank one values with the non-trivial eigenvector being denoted by the blue doubled headed arrows. The insert shows the details of the orientations in the first layering.}
\end{figure}

Our hope was to obtain a similar result as Clark and Milton 
and Clark but for elasticity  in two-dimensional  polycrystals, i.e., characterizing all possible functions $\BC_*(\BC^{(0)})$ of the effective elasticity tensor $\BC_*$ as a function of the  elasticity tensor $\BC^{(0)}$ of the constituent crystal. Specifically, we hoped to show that the function $\BC_*(\BC^{(0)})$ can be approximated to an arbitrarily high degree of approximation by the function for a sequential laminate geometry and such functions are easily characterized through continued fraction expansions. 
Some indication that this might be true comes from a result of  Avellaneda  Cherkaev, Gibiansky, Milton and Rudelson
\cite{Avellaneda:1996:CCP},, who obtained a complete characterization of all possible isotropic $\BC_*$ for a fixed real $\BC^{(0)}>0$ having orthotropic symmetry. They found that 
the effective bulk modulus $\Gk_*$ and effective shear modulus $\Gm_*$ are confined to a rectangular box in the $(\Gk_*,\Gm_*)$
plane and that any point within this rectangle is realized by a hierarchical laminate geometry.  For crystals having more general symmetry
the moduli are confined to a box, with two corners being attained by hierarchical laminate geometries, and numerical results strongly indicating that a third corner is attained by  a hierarchical laminate geometry \cite{Milton:2021:PPE}. The attainability of the fourth corner by  a hierarchical laminate geometry was not established, but seems likely.

What we accomplish is more modest. We obtain a continued fraction approximation to $\BC_*(\BC^{(0)})$ that generalizes the form of the continued fractions associated with
sequential laminates. 

The physical relevance of this result applies to three dimensional materials with a columnar structure with columns perpendicular
to a plane each made from a single crystal rotated in the plane. Provided the single crystal has sufficient symmetry and has the right orientation relative to the direction of each column the three-dimensional elasticity equations decouple into antiplane elasticity equations and the two-dimensional elasticity equations that are the focus of our study (see, for example, Section 2.7 in \cite{Milton:2002:TOC}). Furthermore, if the applied fields are time varying at a frequency $\Go$ and quasistatic (i.e. such that the macroscopic wavelengths and attenuation lengths are much larger than the microstructure) then $\BC^{(0)}=\BC^{(0)}(\Go)$ depends on $\Go$ and may be complex if viscoelasticity is present. The function $\BC_*(\BC^{(0)})$ then gives the dependence of $\BC_*$ on $\Go$: $\BC_*(\Go)=\BC_*(\BC^{(0)}(\Go))$.

\section{Formulating the problem}
%%%%%%%%%%%%%%%%%%%%%%%%%%%%%%%%%%%%%%%%%%%%%%%%%%%%%%%%%%%%%%%%%%%%%%%%%%%%%
\setcounter{equation}{0}
We start by considering the Hilbert space of square integrable  $Q$-periodic $2\times 2$ matrix fields
\beq \BA(\Bx)=\bpm a_{11}(\Bx) &  a_{12}(\Bx) \cr a_{21}(\Bx) &  a_{22}(\Bx) \epm,
\eeq{2.1}
that may be represented by the vector field
\beq \BA(\Bx)=\bpm a_{11}(\Bx) \cr  a_{21}(\Bx) \cr a_{12}(\Bx) \cr  a_{22}(\Bx) \epm.
\eeq{2.2}
The inner product between two fields $\BA(\Bx)$ and  $\BB(\Bx)$ is chosen to be independent of whether we use a matrix or vector representation:
\beqa (\BA,\BB) & = & \int_Q \Tr(\BA(\Bx)\overline{\BB(\Bx)}^T)\,d\Bx \nonum
& = &\int_Q \bpm a_{11}(\Bx) \cr  a_{21}(\Bx) \cr a_{12}(\Bx) \cr  a_{22}(\Bx) \epm
\cdot \bpm \overline{b_{11}(\Bx} \cr  \overline{b_{21}(\Bx)} \cr \overline{b_{12}(\Bx)} \cr  \overline{b_{22}(\Bx)} \epm\,d\Bx \nonum
& = &\int_Q  a_{11}(\Bx) \overline{b_{11}(\Bx)}+a_{12}(\Bx) \overline{b_{12}(\Bx)}+
a_{21}(\Bx) \overline{b_{21}(\Bx)}+a_{22}(\Bx) \overline{b_{22}(\Bx)}\,d\Bx, \nonum &~&
\eeqa{2.2a}
where the overline denotes complex conjugation.  All the subspaces $\BCS$ we consider in this paper will be subspaces of this Hilbert space, equipped with the same norm. Additionally $\Real(\BCS)$ will always be subspace of $\BCS$ and $\BCS=\Real(\BCS)+i\Real(\BCS)$: thus $\BCS$ is the complex extension of
$\Real(\BCS)$ . When we refer to the dimension of $\BCS$ we will mean the dimension of  
$\Real(\BCS)$: a basis of $\Real(\BCS)$ will also serve as a basis for $\BCS$ if we allow complex coefficients in the span. 
\eq{2.2a}.
%The projection onto constant fields, with $\BA(\Bx)$ independent of $\Bx$ will be denoted as %$\BGG_0$. 

The matrix field $\Grad\Bu$ (where we choose to associate the
gradient with the first index is represented by
\beq \Grad\Bu=\bpm \Md u_{1}/\Md x_1 \cr  \Md u_{1}/\Md x_2 \cr \Md u_{2}/\Md x_1 \cr \Md u_{2}/\Md x_2 \epm,
\eeq{2.3}
and we let $\BCE$ denote the space of such fields where $\Bu(\Bx)$ is periodic,
so $\lang\Grad\Bu\rang=0$. We let $\BCJ$ denote the space of current pairs
\beq \bpm \Bj_1 \cr \Bj_2 \epm, \eeq{2.4}
with $\Bj_1$ and $\Bj_2$ both being two component vector fields satisfying
\beq \Div\Bj_1=0,\quad  \Div\Bj_2=0,\quad  \lang\Bj_1\rang=0,\quad \lang\Bj_2\rang=0.
\eeq{2.5}
We introduce two $90^\circ$ rotations:
\beq \BR_\perp=\bpm 0 & 1 & 0 & 0 \cr -1 & 0 & 0 & 0 \cr 0 & 0 & 0 & 1 \cr 0 & 0 & -1 & 0 \epm, \quad
\BCR_\perp=\bpm 0 & 0 & 1 & 0 \cr 0 & 0 & 0 & 1 \cr -1 & 0 & 0 & 0 \cr 0 & -1 & 0 & 0 \epm, \quad
\eeq{2.6}
satisfying
\beq \BR_\perp\BR_\perp=-\BI,\quad \BCR_\perp\BCR_\perp=-\BI,\quad  \BR_\perp^\dagger=\BR_\perp, \quad \BCR_\perp^\dagger=\BCR_\perp,\quad
\BR_\perp\BCR_\perp= \BCR_\perp\BR_\perp.
\eeq{2.6a}
With these a rotation acting on the $2\times 2$ matrix field $\BA$ producing
\beq  \bpm \cos\Gt & \sin\Gt \cr -\sin\Gt & \cos \Gt \epm \bpm a_{11} &  a_{12} \cr a_{21} &  a_{22} \epm \bpm \cos\Gt & -\sin\Gt \cr \sin\Gt & \cos \Gt \epm
\eeq{2.7} 
is represented by first representing the rotation on the left:
\beq \bpm a_{11}\cos\Gt+a_{21}\sin\Gt \cr -a_{11}\sin\Gt+a_{21}\cos\Gt \cr a_{12}\cos\Gt+a_{22}\sin\Gt \cr -a_{12}\sin\Gt+a_{22}\cos\Gt
\epm=(\BI\cos\Gt+\BR_{\perp}\sin\Gt) \bpm a_{11} \cr  a_{21} \cr a_{12} \cr  a_{22} \epm\equiv
\bpm b_{11} \cr  b_{21} \cr b_{12} \cr  b_{22} \epm,
\eeq{2.8}
followed by  representing the rotation on the right:
\beq \bpm b_{11}\cos\Gt+b_{12}\sin\Gt \cr b_{21}\cos\Gt+b_{22}\sin\Gt \cr
-b_{11}\sin\Gt+b_{12}\cos\Gt \cr-b_{21}\sin\Gt+b_{22}\cos\Gt \epm 
=(\BI\cos\Gt+\BCR_{\perp}\sin\Gt) \bpm b_{11} \cr  b_{21} \cr b_{12} \cr  b_{22}
\epm.
\eeq{2.9}
In other words, the full rotation acting on the matrix $\BA$ is represented by
\beq (\BI\cos\Gt+\BCR_{\perp}\sin\Gt)(\BI\cos\Gt+\BR_{\perp}\sin\Gt)
\eeq{2.10}
acting on the vector $\BA$.
The constitutive law is
\beq \BJ(\Bx)=\BL(\Bx)\BE(\Bx),\quad \BJ\in\BCU\oplus\BCJ,\quad \BE\in\BCU\oplus\BCE,
\eeq{2.11}
where, as we have a polycrystal, the $4\times 4$ matrix $\BL(\Bx)$ takes the form
\beq \BL(\Bx) = \BCR(\Bx)\BR(\Bx)\BL^{(0)}\BR^T(\Bx)\BCR^T(\Bx), \eeq{2.12}
in which $\BL^{(0)}$ is the tensor of the pure crystal while $\BCR(\Bx)$ and $\BR(\Bx)$ are the rotations.
\beq \BCR(\Bx)=\BI\cos\Gt(\Bx)+\BCR_{\perp}\sin\Gt(\Bx),\quad \BR(\Bx)=\BI\cos\Gt(\Bx)+\BR_{\perp}\sin\Gt(\Bx).
\eeq{2.14}
In this representation $\BL^{(0)}$ is represented by a $8\times 8$ matrix
\beq \BL^{(0)}=\bpm L^{(0)}_{11} & L^{(0)}_{12} & L^{(0)}_{13} & L^{(0)}_{14} \cr
L^{(0)}_{21} & L^{(0)}_{22} & L^{(0)}_{23} & L^{(0)}_{24} \cr
L^{(0)}_{31} & L^{(0)}_{32} & L^{(0)}_{33} & L^{(0)}_{34} \cr
L^{(0)}_{41} & L^{(0)}_{42} & L^{(0)}_{43} & L^{(0)}_{44} \cr \epm.
\eeq{2.14a}
Here $\BL^{(0)}$  could be the elasticity tensor $\BC^{(0)}$, which is self-adjoint with a
null space consisting of antisymmetric matrices (with $a_{21}=-a_{12}$) and a range consisting of symmetric matrices (with $a_{21}=a_{12}$),
i.e. stresses. Or we could shift $\BC^{(0)}$ by a multiple $c$ of the ``translation'' $\BT$ 
defined by \eq{b.7a} to obtain an equivalent problem with a positive definite symmetric
$\BL^{(0)} =\BC^{(0)} +c\BT$ and effective tensor $\BL_* =\BC_* +c\BT$. 

We define
\beq \Gc(\Bx)=\BCR(\Bx)\BR(\Bx)\Gc^{(0)}\BR^T(\Bx)\BCR^T(\Bx),
\eeq{2.13}
where $\Gc^{(0)}$ is the projection
\beq \Gc^{(0)}=\Bt\otimes\Bt, \quad\quad \Bt=\bpm 1 \cr 0 \cr 0 \cr 0 \epm.
\eeq{2.13aa}
It has the properties that
\beq \Gc\Gc=\Gc,\quad \Gc\BR_\perp\Gc =0,\quad \Gc\BCR_\perp\Gc=0, \text{  and }\Gc\BR_\perp\BCR_\perp\Gc=0. 
\eeq{2.13a}
where the second identity follows from 
\beqa \Gc\BR_\perp\Gc& = & \BCR(\Bx)\BR(\Bx)\Bt\otimes\Bt\BR^T(\Bx)\BCR^T(\Bx)\BR_\perp\BCR(\Bx)\BR(\Bx)\Bt\otimes\Bt\BR^T(\Bx)\BCR^T(\Bx) \nonum
& = & \BCR(\Bx)\BR(\Bx)\Bt(\Bt\cdot\BR_\perp\Bt)\otimes\Bt\BR^T(\Bx)\BCR^T(\Bx)=0,
\eeqa{2.13b}
and similar analysis implies to the third and fourth identities. Another relation follows from
\beq \Bt\otimes\Bt-\BR_\perp\Bt\otimes\Bt\BR_\perp-\BCR_\perp\Bt\otimes\Bt\BCR_\perp+\BCR_\perp\BR_\perp\Bt\otimes\Bt\BR_\perp\BCR_\perp=\BI,
\eeq{o.7}
which upon applying the rotations implies
\beq \Gc-\BR_\perp\Gc\BR_\perp-\BCR_\perp\Gc\BCR_\perp+\BCR_\perp\BR_\perp\Gc\BR_\perp\BCR_\perp=\BI,
\eeq{o.8}
where the four terms on the left project onto four orthogonal subspaces. 
Next note that $\BL^{(0)}$ can be expressed as a linear combination of 16 operators:
\beqa \BL^{(0)} & = & L^{(0)}_{11}\Gc^{(0)}+L^{(0)}_{12}\Gc^{(0)}\BR_\perp+L^{(0)}_{13}\Gc^{(0)}\BCR_\perp+L^{(0)}_{14}\Gc^{(0)}\BR_\perp\BCR_\perp \nonum
& - & L^{(0)}_{21}\BR_\perp\Gc^{(0)} -L^{(0)}_{22}\BR_\perp\Gc^{(0)}\BR_\perp-L^{(0)}_{23}\BR_\perp\Gc^{(0)}\BCR_\perp-L^{(0)}_{24}\BR_\perp\Gc^{(0)}\BR_\perp\BCR_\perp \nonum
& - & L^{(0)}_{31}\BCR_\perp\Gc^{(0)}-L^{(0)}_{32}\BCR_\perp\Gc^{(0)}\BR_\perp-L^{(0)}_{33}\BCR_\perp\Gc^{(0)}\BCR_\perp-L^{(0)}_{34}\BCR_\perp\Gc^{(0)}\BR_\perp\BCR_\perp \nonum
& + & L^{(0)}_{41}\BCR_\perp\BR_\perp\Gc^{(0)}+L^{(0)}_{42}\BCR_\perp\BR_\perp\Gc^{(0)}\BR_\perp+L^{(0)}_{43}\BCR_\perp\BR_\perp\Gc^{(0)}\BCR_\perp
+L^{(0)}_{44}\BCR_\perp\BR_\perp\Gc^{(0)}\BR_\perp\BCR_\perp. \nonum &~&
\eeqa{2.15.0}
Consequently, since $\BR(\Bx)$ and $\BCR(\Bx)$ commute with the operators $\BR_\perp$ and
$\BCR_\perp$ and hence with the rotations $\BCR(\Bx)$ and $\BR(\Bx)$,
$\BL(\Bx)$ can also be expressed  as a linear combination of 16 operators obtained by replacing
$\Gc^{(0)}$ in \eq{2.15.0} with $\Gc$:
\beqa \BL& = & L^{(0)}_{11}\Gc+L^{(0)}_{12}\Gc\BR_\perp+L^{(0)}_{13}\Gc\BCR_\perp+L^{(0)}_{14}\Gc\BR_\perp\BCR_\perp \nonum
& - & L^{(0)}_{21}\BR_\perp\Gc -L^{(0)}_{22}\BR_\perp\Gc\BR_\perp-L^{(0)}_{23}\BR_\perp\Gc\BCR_\perp-L^{(0)}_{24}\BR_\perp\Gc\BR_\perp\BCR_\perp \nonum
& - & L^{(0)}_{31}\BCR_\perp\Gc-L^{(0)}_{32}\BCR_\perp\Gc\BR_\perp-L^{(0)}_{33}\BCR_\perp\Gc\BCR_\perp-L^{(0)}_{34}\BCR_\perp\Gc\BR_\perp\BCR_\perp \nonum
& + & L^{(0)}_{41}\BCR_\perp\BR_\perp\Gc+L^{(0)}_{42}\BCR_\perp\BR_\perp\Gc\BR_\perp+L^{(0)}_{43}\BCR_\perp\BR_\perp\Gc\BCR_\perp
+L^{(0)}_{44}\BCR_\perp\BR_\perp\Gc\BR_\perp\BCR_\perp. \nonum &~&
\eeqa{2.15} 
Hence, as $\BL^{(0)}$ varies, $\BL$ spans a $16$-dimensional space of matrices that is closed under multiplication on the left or right by $\BR_{\perp}$ or $\BCR_{\perp}$. One can see, using \eq{2.6a}, that  multiplying $\BL$ 
on the left or right by $\BR_{\perp}$ or $\BCR_{\perp}$ causes the coefficients $L^{(0)}_{ij}$ to permute positions. 

Note that we are free to make a rotation and redefine
\beq \BGc^{(0)}=\Bt\otimes\Bt, \quad\quad \Bt=\bpm \cos\Gt_0 \cr \sin\Gt_0 \cr \sin\Gt_0 \cr \cos\Gt_0 \epm,
\eeq{2.15a}
where $\Bt$ now corresponds to the symmetric matrix
\beq \bpm \cos\Gt_0 \cr \sin\Gt_0 \cr \sin\Gt_0 \cr \cos\Gt_0 \epm.
\eeq{2.15b}
Correspondingly, $\BGc$ gets replaced by 
\beq \BGc_{\Gt_0}=(\cos\Gt_0\BI+\sin\Gt_0\BR_\perp)(\cos\Gt_0\BI+\sin\Gt_0\BCR_\perp)\BGc(\cos\Gt_0\BI+\sin\Gt_0\BCR_\perp)^T(\cos\Gt_0\BI+\sin\Gt_0\BR_\perp)^T,
\eeq{2.15c} 
and the expansion \eq{2.15} still holds but with associated
adjustments to the coefficients $L^{(0)}_{ij}$. 

Now, in Fourier space, the projection onto $\BCE$ takes the form
\beqa \BGG_1& = & \bpm \frac{\Bk\otimes\Bk}{|\Bk|^2} & 0 \cr  0 & \frac{\Bk\otimes\Bk}{|\Bk|^2} \epm \quad\text{if}\quad \Bk\ne 0 \nonum
& = & 0  \quad\text{if}\quad \Bk= 0,
\eeqa{2.16}
while the projection onto $\BCJ$ takes the form
\beqa \BGG_2& = & \bpm \frac{\Bk_\perp\otimes\Bk_\perp}{|\Bk|^2} & 0 \cr  0 & \frac{\Bk_\perp\otimes\Bk_\perp}{|\Bk|^2} \epm \quad\text{if}\quad \Bk\ne 0 \nonum
& = & 0  \quad\text{if}\quad \Bk= 0,
\eeqa{2.17}
where $\Bk_\perp$ is the $90^\circ$ clockwise rotation of the vector $\Bk$. The projection onto
$\BCU$, the space of constant fields, in Fourier space is
\beqa \BGG_0& = &  0 \quad\text{if}\quad \Bk\ne 0 \nonum
& = & \bpm \BI  & 0 \cr 0 & \BI \epm\quad\text{if}\quad \Bk= 0,
\eeqa{2.17aa}
and $\BGG_0\BA=\lang\BA\rang$ for any field $\BA$. The three spaces $\BCU$, $\BCE$,
$\BCJ$ are mutually orthogonal and
\beq \BGG_0+\BGG_1+\BGG_2=\BI. \eeq{2.17b}
These three projection operators commute with $\BCR_\perp$ while $\BR_\perp\BGG_1=\BGG_2$
and  $\BR_\perp\BGG_0=\BGG_0$. Defining $\BGY$ as the projection
\beq \BGY=\BGY=\BGc-\BR_\perp\BGc\BR_\perp=\BCR(\Bx)\bpm \BI & 0 \cr 0 & 0 \epm \BCR^T(\Bx),
\eeq{2.22aa}
we see that 
\beq \BGY\BGG_0=\BGG_0\BGY, \quad \BGY\BGG_1=\BGG_1\BGY, \quad \BGY\BGG_2=\BGG_2\BGY.
\eeq{2.22aaa}
The projection $\BP$ onto antisymmetric matrix valued fields takes the form
\beq \BP=\bpm 0 & 0 & 0 & 0 \cr 0 & \frac{1}{2} & - \frac{1}{2}& 0 \cr 0 & -\frac{1}{2}& \frac{1}{2}& 0 \cr  0 & 0 & 0 & 0
\epm,
\eeq{2.17a}
which can be written alternatively as
\beq \BP=  - \half\BR_\perp\BGc^{(0)}\BR_\perp+\half\BR_\perp\BGc^{(0)}\BCR_\perp 
 +  \half\BCR_\perp\BGc^{(0)}\BR_\perp-\half\BCR_\perp\BGc^{(0)}\BCR_\perp,
\eeq{2.17bb}
where $\BGc^{(0)}=\Bt\otimes\Bt$. The projection $\BP$ has the properties that
\beq \BP\Bt=0,\quad \BCR_\perp\BR_\perp\BP=\BP\BCR_\perp\BR_\perp=\BP, \quad (\BCR_\perp+\BR_\perp)\BP=\BP(\BCR_\perp+\BR_\perp)=0.
\eeq{2.17c}
These imply that $\BCR(\Bx)\BR(\Bx)$ commutes with $\BP$,
\beqa \BCR(\Bx)\BR(\Bx)\BP& = &\left[\BI\cos^2\Gt(\Bx)+(\BCR_{\perp}+\BR_{\perp})\sin\Gt(\Bx)\cos\Gt(\Bx)+\BCR_\perp\BR_{\perp}\sin^2\Gt(\Bx)\right]\BP \nonum
&=&\BP\left[\BI\cos^2\Gt(\Bx)+(\BCR_{\perp}+\BR_{\perp})\sin\Gt(\Bx)\cos\Gt(\Bx)+\BCR_\perp\BR_{\perp}\sin^2\Gt(\Bx)\right] \nonum
& = &\BP\BCR(\Bx)\BR(\Bx),
\eeqa{2.17d}
as may be expected from the fact that these rotations preserve the symmetry, or antisymmetry, of matrices.

Using this commutivity we see that $\BP=\BCR(\Bx)\BR(\Bx)\BP\BCR^T(\Bx)\BR^T(\Bx)$, and
substituting \eq{2.17bb} into the last expression gives
\beq \BP= - \half\BR_\perp\BGc\BR_\perp+\half\BR_\perp\BGc\BCR_\perp +  \half\BCR_\perp\BGc\BR_\perp-\half\BCR_\perp\BGc\BCR_\perp.
\eeq{2.17e}

\section{The abstract setting}
%%%%%%%%%%%%%%%%%%%%%%%%%%%%%%%%%%%%%%%%%%%%%%%%%%%%%%%%%%%%%%%%%%%%%%%%%%%%%
\setcounter{equation}{0}

Throughout the paper we will successively truncate the Hilbert space $\BCH$ making no changes at any stage to the operators $\Gc$, $\BR_{\perp}$, $\BCR_{\perp}$ and hence $\BL$, and only changing
$\BCU$, $\BCE$, and $\BCJ$ and the operators $\BGG_0$, $\BGG_1$, and $\BGG_2$ that project onto these spaces. Consequently, $\BGG_1$ and $\BGG_2$ will be
no longer given by \eq{2.16} and \eq{2.17} and there is no reason to believe that they remain local operators in Fourier space. As $\BCU$ will change we need to introduce the abstract setting  of effective operators in the theory of composites, called the $Z$-problem. One has a Hilbert space $\BCH$, equipped with an inner product $(\cdot,\cdot)$,
that has a decomposition into three orthogonal subspaces $\BCU$, $\BCE$, and $\BCJ$:
\beq \BCH=\BCU\oplus\BCE\oplus\BCJ. \eeq{h.1}
Given an operator $\BL:\BCH\to\BCH$ one considers the equation
\beq \Bj_0+\Bj=\BL(\Be_0+\Be) \text{     with   } \Bj_0,\Be_0\in \BCU,\quad \Bj\in\BCJ,\quad \Be\in\BCE.
\eeq{1-1}
If for each $\Be_0\in\BCU$ this has a unique solution for $\Bj_0$, $\Bj\in\BCJ$, and $\Be\in\BCE$,
then since $\Bj_0$ depends linearly on $\Be_0$ one may write
\beq \Bj_0=\BL_*\Be_0, \eeq{1-2}
which defines the effective operator $\BL_*:\BCU\to\BCU$. We emphasize that $\BL_*$ is an operator mapping $\BCU$ to $\BCU$. In particular, for our problem, it is not the case that after truncation $\Bj_0(\Bx)=\BL_*\Be_0(\Bx)$ for some $\BL_*$,  possibly dependent on $\Bx$. 
To guarantee existence and uniqueness of the solutions to \eq{1-1}
for all $\Be_0\in\BCU$ it suffices [see, for example, Section 2.4 of \cite{Milton:2016:ETC}] that  $\BL$ be bounded, i.e.
there exists a constant $\Gb>0$  such that for all $\Ba\in\BCH$,
\beq |\BL\Ba|\leq \Gb |\Ba|\quad \text{where}\quad |\Ba|=\sqrt{(\Ba,\Ba)},
\eeq{bed}
and additionally that $\BL$ is $\phi$-coercive for some angle $\phi\in [0,2\pi)$ by which we mean that there exists a constant $\Ga>0$ such that
for all  $\Ba\in\BCH$,
\beq \Real[(e^{i\phi}\BL\Ba,\Ba)]\geq \Ga|\Ba|^2.
\eeq{coer}
For our polycrystal these are satisfied if $\BL_0$ is bounded and the self-adjoint part of the matrix $e^{i\phi}\BL_0$ has a positive definite real part.   Without loss of generality one may set $\phi=0$ by making a rotation of $\BL$ in the complex plane if necessary.

As stated, we will be truncating $\BCH$ making no changes at any stage to the operators $\Gc$, $\BR_{\perp}$, $\BCR_{\perp}$ and hence $\BL$. Therefore, these operators retain their 
properties which we relist here: 
\beq \BGc^2=\BGc,\quad \BGc^\dagger=\BGc, \eeq{o.2}
\beq \BCR_\perp^2=-\BI,\quad \BR_\perp^2=-\BI,\quad \BCR_\perp^\dagger=-\BCR_\perp, \quad \BR_\perp^\dagger=-\BR_\perp, \quad
\BCR_\perp\BR_\perp=\BR_\perp\BCR_\perp,
\eeq{o.3}
\beq \BGc\BR_\perp\BGc =0,\quad \BGc\BCR_\perp\BGc=0, \text{  and }\BGc\BR_\perp\BCR_\perp\BGc=0,
\eeq{o.5}
\beq \BGc-\BR_\perp\BGc\BR_\perp-\BCR_\perp\BGc\BCR_\perp+\BCR_\perp\BR_\perp\BGc\BR_\perp\BCR_\perp=\BI,
\eeq{2.22.1}
At each truncation stage we want the new spaces
$\BCU$, $\BCE$ and $\BCJ$  to be such that the projections $\BGG_0$,
$\BGG_1$ and $\BGG_2$ onto them retain their essential properties:
\beq \BGG_0+\BGG_1+\BGG_2=\BI, \quad \BGG_i\BGG_j=\Gd_{ij}\BGG_i,\quad \BGG_i^\dagger=\BGG_i.
\eeq{o.1}
\beq \BCR_\perp\BGG_i=\BGG_i\BCR_\perp, \text{ for }i=0,1,2, \quad  \BR_\perp\BGG_1=\BGG_2\BR_\perp,\quad  \BR_\perp\BGG_0=\BGG_0\BR_\perp.
\eeq{o.4}
and 
\beq \BGY\BGG_0=\BGG_0\BGY, \quad \BGY\BGG_1=\BGG_1\BGY, \quad \BGY\BGG_2=\BGG_2\BGY.
\eeq{2.22a}
where $\BGY$ is the projection
\beq \BGY=\BGc-\BR_\perp\BGc\BR_\perp=\BI+\BCR_\perp\BGc\BCR_\perp-\BCR_\perp\BR_\perp\BGc\BR_\perp\BCR_\perp, \eeq{2.22b}
in which the last identity follows from \eq{2.22.1}.
From \eq{o.5} $\BGY$ satisfies
\beq \BGY\BGc=\BGc=\BGc\BGY, \quad \BGY(\BCR_\perp\BGc\BCR_\perp)=0,
\eeq{2.22c0}
and additionally,
\beq \BR_{\perp}\BGY=\BR_{\perp}\BGc+\BGc\BR_{\perp}=\BGY\BR_{\perp},\quad
\BCR_{\perp}\BGY=(\BI-\BGY)\BCR_{\perp}. \eeq{2.22!}
The latter implies $\BGY$ and $\BI-\BGY$ project onto spaces of equal dimension. 
We may introduce
\beq \BP= - \half\BR_\perp\BGc\BR_\perp+\half\BR_\perp\BGc\BCR_\perp +  \half\BCR_\perp\BGc\BR_\perp-\half\BCR_\perp\BGc\BCR_\perp,
\eeq{2.22bb}
which according to this definition and \eq{o.5} satisfies
\beq \BP\BGc  =  \BGc\BP=0, \quad \BP^2=\BP, \quad \BP^\dagger=\BP, 
\eeq{2.22c}
the latter two equations implying that $\BP$ is an orthogonal projection, and 
 \beq \BCR_\perp\BR_\perp\BP = \BP\BCR_\perp\BR_\perp=\BP, \quad (\BCR_\perp+\BR_\perp)\BP=\BP(\BCR_\perp+\BR_\perp)=0.
 \eeq{2.22d}
 The first equation is in fact a corollary of the second. Indeed, multiplying the last equation on the left by $\BCR_\perp$ gives
 \beq 0=\BP(\BCR_\perp\BCR_\perp+\BR_\perp\BCR_\perp)=-\BP+\BP\BR_\perp\BCR_\perp, \eeq{2.22e}
 and taking adjoints establishes the first equation in \eq{2.22d}.
 
 The first step in the truncation is done in the appendix. It reduces the Hilbert space to a finite dimensional vector space with only minor perturbations of the operators $\BGG_1$ and $\BGG_2$ and no change to the other operators. We use the same notations for the relevant operators, and relevant subspaces, despite the fact that $\BCE$, $\BCJ$ and the projections $\BGG_1$ and $\BGG_2$ onto them have changed.  The changes in the appendix ensure that $\BL_*(\BL_0)$ remains almost unchanged on the domain of $\BL_0$'s that are of main physical interest. For the remainder of the paper we assume that all spaces are finite dimensional. 
 
 The next truncation step is reducing the dimension of the Hilbert space by $4$. The spaces $\BCE'$ and $\BCJ'$
 that replace  $\BCE$ and $\BCJ$ will each have dimension 2 less than those spaces.
 The new space $\BCU'$, replacing $\BCU$, will still be $4$-dimensional and so the associated effective tensor $\BL'_*$ can be represented by a $4\times 4$ matrix if we choose a basis for $\BCU'$. The truncation needs to be done in such a way that \eq{o.1}, \eq{o.4}, and \eq{2.22a} remain satisfied
 when $\BGG_0$, $\BGG_1$, and $\BGG_2$ are replaced by  $\BGG_0'$, $\BGG_1'$, and $\BGG_2'$ 
 that are defined as the projections onto $\BCU'$, $\BCE'$ and $\BCJ'$. 
 The latter truncation is then repeated giving a chain of effective tensors that are simply linked to  each other. These links provide a continued fraction approximation for the original effective tensor $\BL_*$ of interest.

 \section{Four fields whose span will be first stripped from the truncated space}
 %%%%%%%%%%%%%%%%%%%%%%%%%%%%%%%%%%%%%%%%%%%%%%%%%%%%%%%%%%%%%%%%%%%%%%%%%%%%%
\setcounter{equation}{0}

The strategy we employ is to successively eliminate fields from the vector space $\BCH$ while retaining the essential structure as reflected in the operator relations listed in the abstract setting
of the problem. Then, the idea is to link the effective tensors as these fields are removed.  By iterating the procedure one obtains a series of links that allow one to develop a continued fraction expansion approximation for the effective tensor of a general polycrystal. In this section we  identify those fields whose span will be removed from the vector space $\BCH$ in the next section. 

Observe that
\beq \BGc,\quad -\BCR_\perp\BGc\BCR_\perp, \quad -\BR_\perp\BGc\BR_\perp,\quad  \BCR_\perp\BR_\perp\BGc\BR_\perp\BCR_\perp
\eeq{2.18}
are projections onto mutually orthogonal subspaces that span $\BCH$. Since  $\BCR_\perp\BCE=\BCE$ and $\BCR_\perp\BCJ=\BCJ$ we deduce that $\BCE$ and $\BCJ$ each have even dimension. 
Furthermore, since $\BR_\perp\BCE=\BCJ$ and hence $\BR_\perp\BCJ=\BCE$ we conclude that $\BCE$ and $\BCJ$ have the same dimension that we denote as $2m$. Hence the Hilbert space $\BCH$ has dimension $4m+4$ since $\BCU$ has dimension 4. Now we make the observation which is the crux of our analysis. We first recognize that
\beq (\BGc-\BCR_\perp\BGc\BCR_\perp)(\BCU\oplus\BCJ)\text{  and  }(-\BR_\perp\BGc\BR_\perp+\BCR_\perp\BR_\perp\BGc\BR_\perp\BCR_\perp)(\BCU\oplus\BCE)
\eeq{2.19}
are mutually orthogonal subspaces and so at least one of them has dimension $\leq 2m+2$. Let us begin by considering
the case where the first subspace has dimension $\leq 2m+2$. 
Then as $\BCU\oplus\BCJ$ has dimension
$2m+4$, the key conclusion is that there exists a subspace of at least dimension two comprised of fields $\Bw$ such that
\beq  (\BGc-\BCR_\perp\BGc\BCR_\perp)\Bw=0,\quad \Bw\in \BCU\oplus\BCJ,
\eeq{2.20}
the latter implying that $\BGG_1\Bw=0$.
Applying $\BGY$ to \eq{2.20} and using \eq{2.22a} allows us to conclude to the existence of fields $\Bv=\BGY\Bw$ such that
\beq \BGc\Bv=\BGG_1\Bv=0,\quad \BGY\Bv=-\BR_\perp\BGc\BR_\perp\Bv=\Bv,\eeq{2.23}
or, alternatively, if $\BGY\Bw=0$ then we will see that the field $\Bw$ must be such that
\beq \BGc\Bw=\BGG_1\Bw=0,\quad (\BI-\BGY)\Bw=\Bw. \eeq{2.23a}
To establish this note that $\BGc$ and $-\BR_\perp\BGc\BR_\perp$
project onto orthogonal subspaces, so $\BGY\Bw=0$ implies
\beq \BGc\Bw=0,\quad \BR_\perp\BGc\BR_\perp\Bw=0,\quad \BCR_\perp\BGc\BCR_\perp\Bw=0, \eeq{2.23b}
where the last follows from \eq{2.20}.
In this second scenario \eq{2.23a} we can let $\Bv=\BCR_\perp\Bw$, implying $\Bw=-\BCR_\perp\Bv$ and then
since $\BCR_\perp\BGc\BCR_\perp\Bw=0$ \eq{2.23a} reduces to
\eq{2.23} and we are back at the first scenario. 
%\beq \BCR_\perp\BGc\BCR_\perp\Bv'=\BGG_1\Bv'=0,\quad (\BI-\BGY)\Bv'=\Bv', \eeq{2.23a}
%where the first identity arises from the fact that $\BGc$ and $-\BR_\perp\BGc\BR_\perp$
%project onto orthogonal subspaces, so $\BGY\Bw=0$ implies $\BGc\Bw=0$

%and $\BCR_\perp\BGc\BCR_\perp\Bw=0$.
In the remaining case that the second subspace in \eq{2.19} has dimension $\leq 2m+2$, we infer the existence of a subspace of at least two dimensions comprised of fields $\Br$ such that
\beq
(-\BR_\perp\BGc\BR_\perp+\BCR_\perp\BR_\perp\BGc\BR_\perp
\BCR_\perp)\Br=0,\quad \Br\in \BCU\oplus\BCE,
\eeq{2.25ba}
Multiplying the first equation on the left by $\BR_\perp$
and letting $\Bw=\BR_\perp\Br$ we see that $\Bw$ satisfies 
\eq{2.20}. So again we are back to the first scenario. 

Now, focusing on the first and essentially only scenario, consider the space $\BCU''$ spanned by the four fields
\beq \Bv, \quad \Bv^\perp=\BCR_\perp\Bv, \quad \Bv_{\perp}=\BR_\perp\Bv, \quad
 \Bv^\perp_{\perp}=\BR_\perp\Bv^\perp=\BR_\perp\BCR_\perp\Bv, \eeq{2.30}
and let $\BGG''_0$ denote the projection onto this space. The space is closed under $\BGc$,
since
\beqa 
\BGc\Bv & = & 0, \nonum
\BGc\Bv_{\perp} & = & \Bv_{\perp}, \nonum
\BGc\Bv^\perp_{\perp} & = & \BGc\BCR_\perp\Bv_\perp=\BGc\BCR_\perp\BGc\Bv_\perp=0,  \nonum
\BGc\Bv^\perp & = &-\BGc\BR_\perp\BCR_\perp\Bv_\perp = -\BGc\BR_\perp\BCR_\perp\BGc\Bv_\perp =0,
\eeqa
{2.25a}
where the second equation is implied by \eq{2.23} and the third and fourth equations are implied by it and \eq{o.5}.

So $\BGG''_0$ must commute with
$\BGc$, $\BR_\perp$, and $\BCR_\perp$, as $\BGG''_0\BCH$ is closed under these operations. This implies $\BGG''_0$ commutes with all the 16 operators
entering the expression \eq{2.15} for $\BL$, and therefore with $\BL$ itself. Moreover since $\BGc\Bv_{\perp}=\Bv_{\perp}$ we deduce from
\eq{2.22c} that
\beq \BP\Bv_{\perp}=\BP\BGc\Bv_{\perp}=0, \eeq{2.25b}
implying that $\Bv_{\perp}$ is a symmetric matrix valued field, and due to \eq{2.22d}  $\BR_\perp\BCR_\perp\Bv_{\perp}=\Bv^\perp$ is also a symmetric matrix valued field.  However, there is no reason to
expect that $\Bv=-\BR_{\perp}\Bv_{\perp}$ and $\Bv_\perp^\perp=-\BCR_{\perp}\Bv_{\perp}$ are 
antisymmetric matrix valued fields unless $\Bv_{\perp}(\Bx)$ is diagonal with zero trace for all $\Bx$. 

We also have
\beqa \BGG_1\Bv& = & 0,\nonum
\BGG_1\Bv^\perp & = & \BCR_\perp\BGG_1\Bv=0, \nonum
\BGG_2\Bv_{\perp}& = & \BGG_2\BR_\perp\Bv=\BR_\perp\BGG_1\Bv=0, \nonum
\BGG_2\Bv^\perp_{\perp}& = &\BGG_2\BR_\perp\Bv^\perp=\BR_\perp\BGG_1\Bv^\perp=0,
\eeqa{2.31}
and, using \eq{2.25a} and \eq{o.5}, 
\beqa \BGY\Bv & = & \Bv, \nonum
\BGY\Bv_\perp & = & \Bv_{\perp}-\BR_\perp\BGc\BR_\perp\BGc\Bv_\perp =\Bv_{\perp},\nonum
\BGY\Bv^\perp & = & -\BR_\perp\BGc\BCR_\perp\BGc\Bv_\perp =0,\nonum
\BGY\Bv^\perp_\perp & = & -\BR_\perp\BGc\BR_{\perp}\BCR_\perp\BGc\Bv_\perp =0.
\eeqa{2.31A}
Note that since $\BGc\Bv^\perp=0$ we have 
$\BGc\BCR_{\perp}\Bv=0$ and hence $\Bw=\Bv$ satisfies 
\eq{2.20}. Also $\Bw=-\BCR_{\perp}\Bv$ satisfies \eq{2.20}
and both $\Bv$ and  $-\BCR_{\perp}\Bv$ are independent. 
Consequently, it suffices that there is a one-dimensional subspace
of $\Bv's$ satisfying \eq{2.23} to guarantee that there is a two-dimensional subspace of $\Bw's$ satisfying \eq{2.20}.

\section{Splitting the vector space}
%%%%%%%%%%%%%%%%%%%%%%%%%%%%%%%%%%%%%%%%%%%%%%%%%%%%%%%%%%%%%%%%%%%%%%%%%%%%%
\setcounter{equation}{0}
 In this section we show how the vector space $\BCH$ can be split with one part $\BCU''$ being that spanned by the fields identified in the previous section and the other part being its orthogonal complement $\BCH'$. The vector space $\BCH'$ is such that the relevant operators satisfy the properties in the abstract setting. 
We assume $\Bv$ has been normalized with $\|\Bv\|=1$ so that
\beq  \|\Bv\|=\|\Bv_{\perp}\|=  \|\Bv^\perp\|=\|\Bv^\perp_{\perp}\|=1.
\eeq{4.1}
Then we introduce the subspace $\BCU'=(\BI-\BGG''_0)\BCU$, the projection onto which is 
\beq \BGG_0'=\frac{(\BI-\BGG''_0)\BGG_0(\BI-\BGG''_0)}{1-f},
\eeq{4.3}
where we choose
\beq f=(\BGG_0\Bv,\Bv).\eeq{4.2}
to ensure that $\BGG_0'\BGG_0'=\BGG_0'$.  As $\BCU''$ is closed under the action of the projection $\BGY$ so too will be $\BCU''$.

Now since $\BGG_1\Bv=0$ and  $\BGG_1\Bv^\perp=0$ we deduce that
\beq \BGG_2\Bv=\Bv-\BGG_0\Bv\in\BCJ, \quad  \BGG_2\Bv^\perp=\Bv^\perp-\BGG_0\Bv^\perp\in\BCJ,
\eeq{4.4}
and
\beqa \|\Bv-\BGG_0\Bv\|^2& = &(\Bv-\BGG_0\Bv,\Bv-\BGG_0\Bv)=1-2f+f=1-f, \nonum
\|\Bv^\perp-\BGG_0\Bv^\perp\|^2 & = & (\Bv^\perp-\BGG_0\Bv^\perp,\Bv^\perp-\BGG_0\Bv^\perp)=1-2f+f=1-f.
\nonum &~& 
\eeqa{4.5}
Note that the two fields $\Bv-\BGG_0\Bv$ and $\Bv^\perp-\BGG_0\Bv^\perp$ are orthogonal:
\beq (\Bv^\perp-\BGG_0\Bv^\perp,\Bv-\BGG_0\Bv)=
(\BGG_0\Bv^\perp,\Bv)=(\BGG_0\BCR_\perp\Bv,\BGG_0\Bv)=(\BCR_\perp\BGG_0\Bv,\BGG_0\Bv)=0,
\eeq{4.6}
and span a subspace $\BCJ''\subset\BCJ$ that is closed under $\BGY$ since from \eq{2.31A}
\beq \BGY(\Bv-\BGG_0\Bv)=\Bv-\BGG_0\BGY\Bv=\Bv-\BGG_0\Bv, \quad
 \BGY(\Bv^\perp-\BGG_0\Bv^\perp)=-\BGG_0\BGY\Bv^\perp=0.
 \eeq{4.6a}
We take the orthogonal compliment  of  $\BCJ''$ in our space $\BCJ$ to obtain a new space $\BCJ'$ the projection onto which is
\beq \BGG_2'=\BGG_2-\frac{(\Bv-\BGG_0\Bv)\otimes(\Bv-\BGG_0\Bv)}{1-f}
-\frac{(\Bv^\perp-\BGG_0\Bv^\perp)\otimes(\Bv^\perp-\BGG_0\Bv^\perp)}{1-f}.
\eeq{4.7}
Similarly, the fields
\beq \BGG_1\Bv_\perp=\Bv_\perp-\BGG_0\Bv_\perp\in\BCE, \quad  \BGG_1\Bv^\perp_\perp=\Bv^\perp_\perp-\BGG_0\Bv^\perp_\perp\in\BCE
\eeq{4.8}
span a subspace $\BCE''\subset\BCE$ that is closed under $\BGY$ since from \eq{2.31A}
\beq \BGY(\Bv_\perp-\BGG_0\Bv_\perp)=\Bv_\perp-\BGG_0\BGY\Bv_\perp=\Bv_\perp-\BGG_0\Bv_\perp, \quad
\BGY(\Bv^\perp_\perp-\BGG_0\Bv^\perp_\perp)=-\BGG_0\BGY\Bv^\perp_\perp=0.
\eeq{4.8A}
By  taking the orthogonal compliment  of  $\BCE''$ in our space $\BCJ$ we obtain a new space $\BCJ'$ the projection onto which is
\beq \BGG_1'=\BGG_1-\frac{(\Bv_\perp-\BGG_0\Bv_\perp)\otimes(\Bv_\perp-\BGG_0\Bv_\perp)}{1-f}
-\frac{(\Bv^\perp_\perp-\BGG_0\Bv^\perp_\perp)\otimes(\Bv^\perp_\perp-\BGG_0\Bv^\perp_\perp)}{1-f}.
\eeq{4.9}
From \eq{4.4} and \eq{4.8} we get the alternative expressions:
\beqa \BGG_2'& = & \BGG_2-\frac{\BGG_2\Bv\otimes\Bv\BGG_2}{1-f}
-\frac{\BGG_2\Bv^\perp\otimes\Bv^\perp\BGG_2}{1-f}, \nonum
 \BGG_1'& = & \BGG_1-\frac{\BGG_1\Bv_\perp\otimes\Bv_\perp\BGG_1}{1-f}
 -\frac{\BGG_1\Bv^\perp_\perp\otimes\Bv^\perp_\perp\BGG_1}{1-f}.
 \eeqa{4.10}
 Additionally, we define  $\BCH'$ to be the orthogonal complement
 of $\BCU''$ in the space $\BCH$. Then $\BCU'=(\BI-\BGG''_0)\BCU$ consists of those fields in $\BCH'$ that are orthogonal to both $\BCE'$ and $\BCJ'$. As $\BCE$ and $\BCE''$ are closed under the projection $\BGY$, so too will be $\BCE'$, and similarly as $\BCJ$ and $\BCJ''$ are closed under the projection $\BGY$, so too will be $\BCJ'$. In other words \eq{2.22a} holds with $\BGG_i$ replaced by 
 $\BGG_i'$:
 \beq \BGY\BGG_0'=\BGG_0'\BGY, \quad \BGY\BGG_1'=\BGG_1'\BGY, \quad \BGY\BGG_2'=\BGG_2'\BGY.
 \eeq{4.10A}

 \section{Linking effective tensors}
 %%%%%%%%%%%%%%%%%%%%%%%%%%%%%%%%%%%%%%%%%%%%%%%%%%%%%%%%%%%%%%%%%%%%%%%%%%%%%
\setcounter{equation}{0}
In this section we obtain a link between the effective tensor $\BL_*$ associated with $\BCH$ 
and the effective tensor $\BL_*'$ associated with $\BCH'$. As we shall see, the problem of solving for the fields that define $\BL_*$ splits into the problem of solving for the fields that define $\BL_*'$ and an auxillary problem that links the effective tensors.

Our space $\BCE$ has the splitting $\BCE=\BCE'\oplus\BCE''$ while $\BCJ$ has the splitting $\BCJ=\BCJ'\oplus\BCJ''$ . The space $\BCV\equiv\BCU\oplus\BCE''\oplus\BCJ''$ can be decomposed as 
% \beq \BCV=\BCU'\oplus\BCU''=\BCU\oplus\BCU_\perp, \text{   where   }\BCU_\perp=\BCE''\oplus\BCJ''.
%\eeq{5.1}
\beq \BCV'=\BCU\oplus\BCU_\perp=\BCU'\oplus\BCU'', \text{   where     }\quad\BCU_\perp=\BCE''\oplus\BCJ'',
\eeq{5.1}
and $\BCU'$ is defined as the orthogonal complement of $\BCU''$ in the space $\BCV$.
We are interested in solving
\beq \Bj_0+\Bj=\BL(\Be_0+\Be), \quad\text{     where   }\,\, \Bj_0,\,\Be_0\in \BCU,\quad \Bj\in\BCJ,\quad \Be\in\BCE,
\eeq{5.2}
in which $\Bj$ and $\Be$ can be decomposed as 
\beq \Bj=\Bj'+\Bj'',\quad  \Be=\Be'+\Be'',\quad \text{   with   }\,\, \Bj'\in \BCJ', \,\Bj''\in \BCJ'',\quad \Be'\in \BCE', \,\Be''\in\BCE'',
\eeq{5.2a}
where $\Bj''$ and $\Be''$ can be further decomposed as 
\beq 
\Bj''=\underline{\Bj'}+\underline{\Bj''},\quad \Be''=\underline{\Be'}+\underline{\Be''}\,\,\text{     with   }  \underline{\Be''},\,\underline{\Bj''}\in\BCU'', \quad \underline{\Bj'}, \,\underline{\Be'}\in\BCU',
\eeq{5.3}
while $\Bj_0$ and $\Be_0$ can be decomposed as
\beq \Bj_0=\Bj'_0+\Bj''_0,\quad  \Be_0=\Be'_0+\Be''_0, \text{     with   } \Bj'_0,\,\Be'_0\in\BCU',\quad\Bj_0'',\,\Be_0''\in \BCU''. 
\eeq{5.3a}
By the definition of $\BL_*$ we have
\beq \Bj_0=\BL_*\Be_0. \eeq{5.4}
%Now $\Be''$ and $\Bj"$ are in $\BCU_\perp$ so we may write
%\beq \Bj_0+\Bj"=\Bu'_j+\Bu"_j,\quad  \Be_0+\Be"=\Bu'_e+\Bu"_e,\text{  with  }\Bu'_j,\Bu'_e \in\BCU',\quad \Bu''_j,\Bu''_e %\in\BCU''
%\eeq{5.5}
Since $\BL$ commutes with $\BGG''_0$ the  relation in \eq{5.2} splits into two equations:
\beq \Bj''_0+\underline{\Bj''}=\BL(\Be''_0+\underline{\Be''}), \quad  \Bj_0'+\underline{\Bj'}+\Bj'=\BL(\Be_0'+\underline{\Be'}+\Be'').
\eeq{5.6}
The second  equation, in the Hilbert space $\BCH'=\BCU'\oplus\BCJ'\oplus\BCE'$ has
\beq \Bj_0'+\underline{\Bj'},\Be_0'+\underline{\Be'}\in \BCU',\quad \Bj'\in\BCJ',\quad\Be'\in \BCE', \eeq{5.7}
and so there exists an associated effective tensor $\BL_*'$ such that
\beq \Bj_0'+\underline{\Bj}'=\BL_*'(\Be_0'+\underline{\Be'}).
\eeq{5.8}
Here $\BL_*'$ is an operator mapping $\BCU'$ to  $\BCU'$ that can be represented as a matrix if
we introduce a basis for $\BCU'$. 
Together with the first equation in \eq{5.6} implies
\beq \Bj''_0+\underline{\Bj''}+\Bj_0'+\underline{\Bj'}=\BL'(\Be_0''+\underline{\Be''}+\Be_0'+\underline{\Be'}), \eeq{5.9}
where
\beq \BL'=\BL\BGG''_0 +\BL_*'(\BI-\BGG''_0). \eeq{5.10}
We can rewrite \eq{5.9} as
\beq \Bj_0+\Bj''=\BL'(\Be_0+\Be'') \text{  with  }
 \Bj_0,\,\Be_0\in \BCU,\quad \Bj''\in\BCJ'',\quad \Be''\in\BCE''.
\eeq{5.11}
and the associated effective tensor $\BL_*$, relating $\Bj_0$ with $\Be_0$ must be exactly the same as in \eq{5.4}. In other words,
$\BL_*$ is not only the effective tensor associated with $\BL$, but also it is the effective tensor  associated with $\BL'$, that itself depends on $\BL$ and the effective tensor $\BL_*'$ associated with $\BCH'$. 

%$\overline{\BCR}_\perp^{(2)}$ 
%first note that $\BGc\Bv=0$. Next observe
%that
%\beq \BCR_\perp(\BI-\BGY)=\bpm 0 & \BI \cr -\BI & 0 \epm\bpm  0 & 0 \cr 0 & \BI \epm= \bpm 0 & \BI \cr 0 & 0 \epm=\BGY\BCR_\perp
%\eeq{2.25}
%which implies
%\beq \BGc\BCR_\perp\Bv=\BGc\BCR_\perp\BGY\Bv=
%and
%also $\BGc\BCR\Bv=0$ since
%\beq \BGc\BCR\Bv

 \section{The continued fraction expansion approximation for general polycrystals}
%%%%%%%%%%%%%%%%%%%%%%%%%%%%%%%%%%%%%%%%%%%%%%%%%%%%%%%%%%%%%%%%%%%%%%%%%%%%%
\setcounter{equation}{0}

Here we find representations for the various operators needed to solve \eq{5.11} with $\BL'$ being given by \eq{5.10}. This gives $\BL_*$ in terms of $\BL_*'$ and iteration of the procedure results in a continued fraction expansion approximation for $\BL_*$ for general polycrystals.

 The fields $\Bv$, $\Bv^{\perp}$, $\Bv_\perp$, and $\Bv^\perp_{\perp}$ naturally form an orthonormal basis for $\BCU''$ while the fields
 \beqa \Bz& = & \frac{(\BGG_0-f\BI)\Bv}{\sqrt{f(1-f)}},\quad \Bz^\perp=\frac{(\BGG_0-f\BI)\Bv^\perp}{\sqrt{f(1-f)}},\nonum 
 \Bz_\perp& = & \frac{(\BGG_0-f\BI)\Bv_\perp}{\sqrt{f(1-f)}},
 \quad\Bz_\perp^\perp=\frac{(\BGG_0-f\BI)\Bv_\perp^\perp}{\sqrt{f(1-f)}}
 \eeqa{6.1}
 form an associated basis for the orthogonal space $\BCU'$. The two fields
\beq \Bq=\frac{\Bv-\BGG_0\Bv}{\sqrt{1-f}}, \quad \Bq^\perp=\frac{\Bv^\perp-\BGG_0\Bv^\perp}{\sqrt{1-f}} \eeq{6.2}
  form a basis for $\BCJ''$, while
 \beq \Bq_\perp=\frac{\Bv_\perp-\BGG_0\Bv_\perp}{\sqrt{1-f}}, \quad \Bq^\perp_\perp=\frac{\Bv^\perp_\perp-\BGG_0\Bv^\perp_\perp}{\sqrt{1-f}} \eeq{6.3}
 form a basis for $\BCE''$. The associated orthonormal basis for $\BCU$ consists of the four fields
 \beq  \Bh=\frac{\BGG_0\Bv}{\sqrt{f}},\quad\Bh^\perp=\frac{\BGG_0\Bv^{\perp}}{\sqrt{f}},
 \quad\Bh_\perp=\frac{\BGG_0\Bv_\perp}{\sqrt{f}},\quad \Bh_\perp^\perp=\frac{\BGG_0\Bv^\perp_{\perp}}{\sqrt{f}}.
 \eeq{6.4}
 One basis field for the subspace $\BGc\BCH''$ is $\Bv_\perp$. In solving \eq{5.11} a second basis field for $\BGc\BCH''$ can be chosen  to be  $(\BGG_0-f\BI)\Bv_\perp/\sqrt{f(1-f)}\in \BCU'$
 with associated coefficients $L_{*ij}'$ of $\BL'_*$ determined by this  choice. 
 
From \eq{6.1} we deduce that
\beq \BGG_0\Bv=f\Bv+\sqrt{f(1-f)}\Bz,\quad \BGG_0\Bz=\frac{(1-f)\BGG_0\Bv}{\sqrt{f(1-f)}}=(1-f)\Bz+\sqrt{f(1-f)}\Bv,
\eeq{6.4a}	
with similar expressions for the action of $\BGG_0$ on the other basis elements of $\BCU''$ and $\BCU'$. Thus, relative to the basis $\Bv,\Bv^\perp,\Bv_\perp, \Bv^\perp_\perp,
\Bz,\Bz^\perp,\Bz_\perp,\Bz^\perp_\perp$ the operator $\BGG_0$ is represented by the $8\times 8$ matrix
\beq \BGG_0 =\bpm f\BI & \sqrt{f(1-f)}\BI \cr \sqrt{f(1-f)}\BI & (1-f)\BI \epm ,
 \eeq{6.4b}
 where here $\BI$ is the $4\times 4$ identity matrix. 
  Using the fact that $\BGG_0+\BGG_1+\BGG_2=\BI$ and that $\BGG_1\Bv=\BGG_1\Bv^\perp=0$ and $\BGG_1\Bz=\BGG_1\Bz^\perp=0$ (the latter being implied by \eq{6.1} )  we get the 
  $8\times 8$ matrix representations for $\BGG_1$ and $\BGG_2$:
\beqa  
\BGG_1  & = &  \bpm 0 & 0 & 0  & 0\cr
0 & (1-f)\BI & 0  & -\sqrt{f(1-f)}\BI   \cr
0 &  0   & 0 & 0 \cr
0 & -\sqrt{f(1-f)}\BI  & 0 & f\BI \epm,\nonum
\BGG_2 & = & \bpm (1-f)\BI & 0  & -\sqrt{f(1-f)}\BI & 0 \cr 
0 & 0  &0  & 0 \cr
 -\sqrt{f(1-f)}\BI  & 0 & f\BI &0 \cr
0 & 0 & 0 & 0 \epm,
 \eeqa{6.4c}
 where here $\BI$ is the $2\times 2$ identity matrix.
 The operators $\BR_{\perp}$ and $\BCR_\perp$ are clearly represented by the $8\times 8$ matrices
 \beqa\BR_{\perp} & = & \bpm \dund{\BR}_\perp & 0 \cr 0 & \dund{\BR}_\perp\epm\text{    with   }
\dund{\BR}_\perp =\bpm 0 &  0 & -1 & 0\cr
 0  & 0  & 0  & -1 \cr
 1 & 0 & 0 &  0\cr
 0 & 1  & 0  & 0 \epm, \nonum
 \BCR_{\perp} & = &\bpm \dund{\BCR}_\perp & 0 \cr 0 & \dund{\BCR}_\perp\epm\text{    with   }
 \dund{\BCR}_\perp =\bpm 0 & -1 & 0  & 0\cr
 1  & 0  &0  & 0 \cr
 0 & 0 & 0 & -1 \cr
 0 & 0  & 1  & 0 \epm.
  \eeqa{6.4e}
 
 Since $\BCU''$ is closed under $\BGc$ so must be $\BCU'$. In other words, $(\BGG_0-f\BI)/\sqrt{f(1-f)}$  being the projection onto $\BCU''$ must commute with $\BGc$. It follows that
 \beq \BGc\Bz=\BGc\Bv=0,\quad \BGc\Bz^\perp=\BGc\Bv^\perp=0,\quad 
 \BGc\Bz_\perp=\Bz_\perp,\quad \BGc\Bv_\perp=\Bv_\perp, \quad \BGc\Bz_\perp^\perp=\BGc\Bv_\perp^\perp=0,
\eeq{6.4f}
implying that $\BGc$ is represented by the $8\times 8$ matrix
\beq \BGc=\bpm \dund{\BGc} & 0 \cr 0 & \dund{\BGc} \epm \text{    with   }\dund{\BGc}=\bpm 0 & 0 & 0  & 0\cr
 0  & 0  &0  & 0 \cr
 0 & 0 & 1 & 0 \cr
 0 & 0  & 0  & 0 \epm.
 \eeq{6.4g}
 The operator $\BL'$ takes the block form
 \beq  \BL'=\bpm \BL & 0 \cr 0 & \BL_*' \epm,
 \eeq{6.4ha}
 where the $4\times 4$ matrix representing the action of $\BL$ in this basis is given by \eq{2.15} with $\BGc$, $\BR_\perp$, and $\BCR_\perp$ replaced by  $\dund{\BGc}$, $\dund{\BR}_\perp$ ,
 and $\dund{\BCR}_\perp$.

 There are many formulae giving the effective tensor $\BL_*$. One is
 \beq \BL_*=\widetilde{\BGG}_0[(\underline{\BGG_0+\BGG_1})(\BL')^{-1}(\underline{\BGG_0+\BGG_1})^T]^{-1}\widetilde{\BGG}_0^T,
 \eeq{11.1}
 where, according to \eq{5.10}, the operator $(\BL')^{-1}$ takes the form
 \beq   (\BL')^{-1}=\bpm \BL^{-1} & 0 \cr 0 & [\BL_*']^{-1}\epm,
 \eeq{11.3}
 and the $6\times 8$ matrix
 \beq \underline{\BGG_0+\BGG_1}
 =\bpm \sqrt{f}\BI &  0 & \sqrt{1-f}\BI  &0\cr
 0 & \sqrt{f}\BI &  0 & \sqrt{1-f}\BI \cr
 0  & \sqrt{1-f}\BI &  0 & \sqrt{f}\BI \epm
 \eeq{11.2}
 represents the projection $\BGG_0+\BGG_1$ as a map from the basis
 $\Bv$, $\Bv^\perp$, $\Bv_\perp$, $\Bv^\perp_\perp$,
 $\Bz$, $\Bz^\perp$, $\Bz_\perp$, and $\Bz^\perp_\perp$ to the basis  $\Bh$, $\Bh^{\perp}$, $\Bh_\perp$, $\Bh^\perp_{\perp}$, $\Bq_\perp$, and $\Bq_\perp^\perp$ 
 while the $4\times 6$  matrix
 \beq  \widetilde{\BGG}_0 = \bpm \BI & 0 & 0 \cr
 0  & \BI & 0 \epm
 \eeq{11.4a}
 represents the projection $\BGG_0$ as a map from the basis
 $\Bh$, $\Bh^{\perp}$, $\Bh_\perp$, $\Bh^\perp_{\perp}$, $\Bq_\perp$, and $\Bq_\perp^\perp$ 
 to the basis  $\Bh$, $\Bh^{\perp}$, $\Bh_\perp$, and $\Bh^\perp_{\perp}$.
 It is evident from the matrices representing the various operators that the relation 
 $\BL_*(\BL_*')$  between the matrix representing the effective tensor $\BL_*$ and the matrix representing $\BL_*'$ in these bases
 only involves $f$ and the moduli $\BL^{(0)}_{ij}$ of the pure crystal.
 
 By iterating the procedure one gets
  \beq   \BL_*^{(j)}=\widetilde{\BGG}_0[(\underline{\BGG_0+\BGG_1})^{(j)}[\BL'^{(j-1)}]^{-1}(\underline{\BGG_0+\BGG_1})^{(j)T}]^{-1}\widetilde{\BGG}_0^T,
 \eeq{11.1aa}
 where
 \beq   \BL'^{(j-1)}=\bpm \BL^{-1} & 0 \cr 0 & [\BL_*'^{(j-1)}]^{-1}\epm,
 \eeq{11.3aa}
 and $(\underline{\BGG_0+\BGG_1})^{(j)}$ is obtained from $\underline{\BGG_0+\BGG_1}$
 by replacing $f$ with $f^{(j)}$ in \eq{11.2}.
 Our representation is such that the matrix $\BL$ entering \eq{11.3aa} is independent of $j$:  it is given by \eq{2.15} with $\BGc$, $\BR_\perp$, and $\BCR_\perp$ replaced by  $\dund{\BGc}$, $\dund{\BR}_\perp$ ,
and $\dund{\BCR}_\perp$ as defined in \eq{6.4g} and \eq{6.4e}. 
Note that $\BL_*'^{(j-1)}$ and $\BL_*^{(j-1)}$ are both representations of the same effective
 operator, but with respect to different bases. To convert the matrix $\BL_*^{(j-1)}$ to the matrix $\BL_*'^{(j-1)}$ we need to make a change of basis giving
\beq   \BL_*'^{(j-1)}=\BQ^{(j-1)}\BL_*^{(j-1)}\BQ^{(j-1)T} \text{   where  }\BQ^{(j-1)}\BQ^{(j-1)T} =\BI.
\eeq{6.4j}
 Repeated substitution of \eq{6.4j} in \eq{6.4} $m$ times gives a continued fraction expansion 
 beginning with $\BL_*^{(m)}=\BL_*$ and terminating with 
 $\BL_*^{(0)}=\BQ^{(0)}\BL^{(0)}\BQ^{(0)T}$.  (Note that at each stage we reduce the dimension of the space by $4$ beginning with a $4m+4$ dimensional space and ending with a $4$ dimensional space). 
 Some flexibility is added by replacing $\BGc$ at each stage $j$ with
 \beq \BGc_{\phi_j}= \BCR_{\phi_j}\BR_{\phi_j}\BGc\BR_{\phi_j}^T\BCR_{\phi_j}^T,
 \eeq{6.4k}
 where 
 \beq \BCR_{\phi_j}=\BI\cos\phi_j+\BCR_{\perp}\sin\phi_j,\quad \BR_{\phi_j}=\BI\cos\phi_j+\BR_{\perp}\sin\phi_j,
 \eeq{6.4l}
 and we are free to choose the angles $\phi_j$.
 It has the same properties as $\BGc$ and we conclude to the existence of fields $\Bv_{\perp}^{(j)}$, dependent on $\phi_j$,
 such that
 \beq \BGc_{\phi_j}\Bv_\perp^{(j)}=\Bv_\perp^{(j)},\quad \BGG_1^{(j)}\Bv_\perp^{(j)}=0, \quad 
 \BGY^{(j)}\Bv_\perp^{(j)}=\Bv_\perp^{(j)},
  \eeq{6.4m}
  where
  \beq \BGY^{(j)}=\BGc_{\phi_j}-\BR_{\perp}\BGc_{\phi_j}\BR_{\perp}.  \eeq{6.4n}
  The associated fields $\Bv^{(j)}$, $\Bv^{\perp (j)}$, $\Bv_\perp^{\perp (j)}$, $\Bz^{(j)}$, $\Bz^{\perp (j)}$, 
  $\Bz_{\perp (j)}$,  and $\Bz_\perp^{\perp (j)}$,  defined analogously to \eq{2.30} satisfy the expected properties. 
 With these extra degrees of freedom $f_j$ and $\BQ^{(j)}$ will depend on our choice of  the angles $\phi_j$. Also, the $\BL$ entering \eq{11.3aa} will depend on $j$ as when $\BGc$ is replaced with
 $\BGc_{\phi_j}$ the coefficients in the expression \eq{2.15} will be accordingly changed. 
 
 There are still many alternative continued fractions. In particular we are free to use other formulae for the effective tensor
 $\BL_*$ in terms of $\BL_*'$. One such formula is
  \beq \BL_*=\Gs_0\BGG_0-\Gs_0\dund{\BGG}_0(\BS'-\BGG_1)^{-1}
  \dund{\BGG}_0^T, \text{     where   } \BS=\Gs_0(\Gs_0\BI-\BL')^{-1},
 \eeq{6.4h}
 and the  $4\times 8$ matrix
 \beq \dund{\BGG}_0= \bpm \sqrt{f}\BI & \sqrt{1-f}\BI \epm ,\eeq{6.4ba}
 where $\BI$ is the $4\times 4$ identity matrix, represents the projection $\BGG_0$ as a mapping from the basis 
 $\Bv$, $\Bv^\perp$, $\Bv_\perp$, $\Bv^\perp_\perp$,
 $\Bz$, $\Bz^\perp$, $\Bz_\perp$, and $\Bz^\perp_\perp$ to the basis  $\Bh$, $\Bh^{\perp}$, $\Bh_\perp$, $\Bh^\perp_{\perp}$. One may verify that $\dund{\BGG}_0^T\dund{\BGG}_0$ is the matrix on the right hand side of \eq{6.4b} while  $\dund{\BGG}_0\dund{\BGG}_0^T=\BI$, as expected. 
 The result is independent of $\Gs_0$, but generally it should be chosen so that the matrices one wants to invert are non-singular. Here $\BGG_0$ and $\BGG_1$
 are given by \eq{6.4ba} and \eq{6.4c}, while $\BL'$ is given by \eq{6.4ha}.
 
 \section{The possible correspondence  with  sequential laminates}
 %%%%%%%%%%%%%%%%%%%%%%%%%%%%%%%%%%%%%%%%%%%%%%%%%%%%%%%%%%%%%%%%%%%%%%%%%%%%%
\setcounter{equation}{0}

Now consider a laminate geometry with the pure crystal, with its orientation to be determined,
being laminated with a material having tensor $\BL'_*$, in proportions $f$ and $1-f$ respectively.
The resulting effective tensor is $\BL_*$. The space $\BCU$, as usual, is the space of constant fields and $\BGG_0$ is the projection onto it. We define $\BCU''$ as the space of fields that are
 zero in the material having tensor  $\BL'_*$ while $\BCU'$ is the space of fields that are zero in the layers of pure crystal material. 
 The layer interfaces are perpendicular to a unit vector $\Bn$ called the direction of lamination. We aim to set up, if possible, a correspondence with the fields in $\BCH''$ for
 an arbitrary polycrystal. To simplify notation we use the same symbols to denote the corresponding subspaces and corresponding fields.
 We let $\BCU''$ denote the subspace of fields which are
 non-zero only inside the pure crystal layers, and we define $\BGG_0''$ as the projection onto this subspace. Letting $\Bn_\perp=\BR_\perp\Bn$, one field in $\BCU''$ is
 \beqa \Bv^\perp & = & \frac{\Bn_\perp\otimes\Bn_\perp}{\sqrt{f}}\quad\quad\text{in the pure crystal phase} \nonum
                 & = & 0\quad\quad \text{elsewhere},
                 \eeqa{6.5}
                 and another is $\Bv=-\BCR\Bv^\perp$. Both these fields lie in $\BCU\oplus\BCJ$ since $\Div\Bv^\perp=\Div\Bv=0$.  We next define the fields
                 \beqa  \Bv_\perp =\BR_\perp\Bv & = & \frac{\Bn\otimes\Bn}{\sqrt{f}}\quad\quad\text{in the pure crystal phase} \nonum
                 & = & 0\quad\quad \text{elsewhere},
                 \eeqa{6.6}
                 and  $\Bv_\perp^\perp=\BR_\perp\Bv^\perp$.
These both lie in $\BCU\oplus\BCE$, since $\Curl\Bv_\perp=\Curl\Bv_\perp^\perp= 0$. 
Together with $\Bv$ and $\Bv^\perp$ they satisfy \eq{2.30}. As before, we take $\Bv$, $\Bv^\perp$,
$\Bv_\perp$ and $\Bv_\perp^\perp$ as our basis for $\BCU''$. 
Next, we can define $\BGc\BCH''$ as the span of
                 the fields $\Bv_\perp$ and  $(\BGG_0-f\BI)\Bv_\perp/\sqrt{f(1-f)}$ and take these two fields as a basis for  $\BGc\BCH''$. Then \eq{2.25a} and \eq{2.25b} are satisfied. An example of the corresponding sequential laminate is shown in Figure 3(a)
                 \begin{figure}[!ht]
                 	\includegraphics[width=0.9\textwidth]{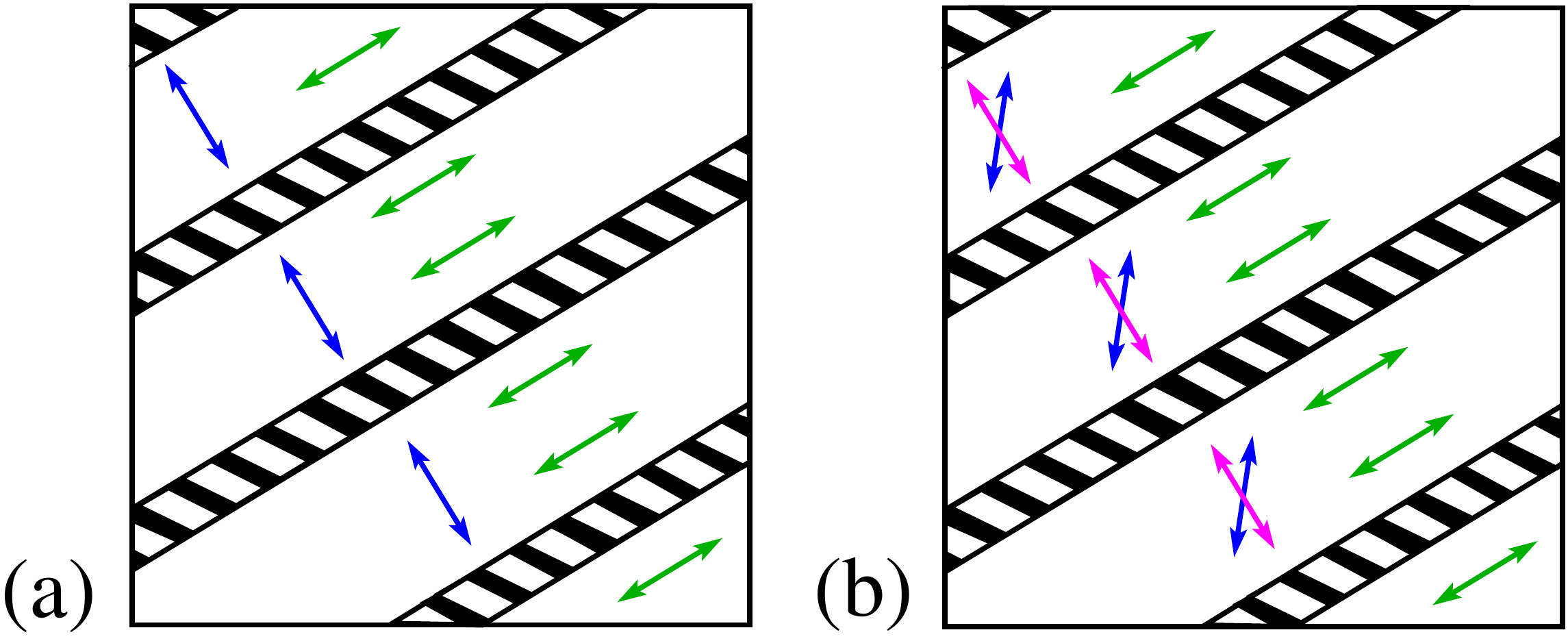}
                 	\caption{(a)  An example of a field $\Bv^\perp$ in a sequential laminate which is nonzero only in the last layers and takes a symmetric rank one value $\Bn_\perp\otimes\Bn_{\perp}/\sqrt{f}$ there.
                 		Here $\Bn$ is the direction of the last lamination, and  $\Bn_\perp=\BR_{\perp}\Bn$. The double headed green arrows denote the non-trivial unit eigenvectors $\pm \Bn_\perp$ of   $\Bv^\perp$ in those last layers, and we see that $\Div\Bv^\perp=0$ implying $\BGG_1\Bv^\perp=0$. The blue double headed blue arrows denote the non-trivial unit eigenvectors $\pm \Bn$ of   $\Gc$ in the last layers, ensuring that $\BGc\Bv^\perp=0$. (b) More general crystal orientations, again with the blue double headed arrows denoting  the non-trivial unit eigenvectors of   $\Gc$, are obtained by looking
                 		for an angle $\Gf$ such that $\Bv^\perp= \Bn_\perp\otimes\Bn_{\perp}/\sqrt{f}$ satisfies
                 		$\BGc_\phi\Bv^\perp=0$. Thus  the non-trivial unit eigenvectors of $\Gc_\phi$ in the last layers are 
                 		$\pm\Bn$, denoted by the magneta double headed arrows, and we have  $\BGG_1\Bv^\perp=0$
                 		and $\BGc_\phi\Bv_{\perp}=0$.}
                 \end{figure}
                 
                 The orientation of the pure crystal
                 needs to be taken so its coefficients $L^{(0)}_{ij}$ agree with those associated with $\BL\BGG''_0$ in \eq{5.10} with respect to the basis of $\BCU''$ taken to be
                 $\Bv$, $\Bv^{\perp}$, $\Bv_\perp$, and $\Bv^\perp_{\perp}$ in both cases. The coefficients  $L^{(0)}_{ij}$ and  $L_{*ij}'$ of $\BL^{(0)}$ and $\BL'_*$ are then fixed. As in
                 the general polycrystal case, we choose the fields \eq{6.1} as a basis for $\BCU'$. The fields in \eq{6.2} are taken as a basis for the subspace spanned by them,
                 defined to be $\BCJ''$. Similarly, the fields in \eq{6.3} are taken as a basis for the subspace spanned by them, defined to be $\BCE''$. Then the action of $\BGG_1$ and
                 $\BGG_2$ on $\BCH''$ is the same as the action of $\BGG''_1$ and $\BGG''_2$ on $\BCH''$, each defined as the projections onto $\BCE''$ and $\BCJ''$. For the sequential laminate we iterate this procedure and the
                 relations \eq{11.1aa} and \eq{11.3aa} will still hold. Again
$\BL_*'^{(j-1)}$ and $\BL_*^{(j)}$ are both representations of the same effective
operator, but with respect to different basis fields. The first is respect to a basis of the form
\beqa  \Bz^{(j)}& = & \frac{(\BGG_0^{(j)}-f_{j}\BI)\Bv^{(j)}}{\sqrt{f_{j}(1-f_{j})}},\quad
 \Bz^{\perp(j)} =  \frac{(\BGG_0^{(j)}-f_j\BI)\Bv^{\perp(j)}}{\sqrt{f_j(1-f_j)}}, \nonum
\Bz_\perp^{(j)} & = &\frac{(\BGG_0^{(j)}-f_j\BI)\Bv_\perp^{(j)}}{\sqrt{f_j(1-f_j)}},\quad
 \quad\Bz_\perp^{\perp(j)}=\frac{(\BGG_0^{(j)}-f_j\BI)\Bv_\perp^{\perp(j)}}{\sqrt{f_j(1-f_j)}},
 \eeqa{6.7a}
 while the second is respect to a basis of the form
 \beqa  \Bh^{(j-1)}& = & \frac{\BGG_0^{(j-1)}\Bv^{(j-1)}}{\sqrt{f_{j-1}}},\quad\Bh^{\perp(j-1)}=\frac{\BGG_0^{(j-1)}\Bv^{\perp(j-1)}}{\sqrt{f_{j-1}}}, \nonum
\quad\Bh_\perp^{(j-1)}& = & \frac{\BGG_0^{(j-1)}\Bv_\perp^{(j-1)}}{\sqrt{f_{j-1}}},\quad \Bh_\perp^{\perp(j-1)}=\frac{\BGG_0^{(j-1)}\Bv_\perp^{\perp(j-1)}}{\sqrt{f_{j-1}}},
\eeqa{6.8}
where each basis spans the same space, $\BCU'^{(j)}=\BCU^{(j-1)}$. Now because $\Bv^{\perp(j)}$ is symmetric and rank one for all $j$, having unit norm, so too will be
$\Bz^{\perp(j)}$ and $\Bh^{\perp(j-1)}$. So there will be rotations,
\beq \BCR_{j-1}=\BI\cos\Gt_{j-1}+\BCR_{\perp}\sin\Gt_{j-1},\quad \BR_{j-1}=\BI\cos\Gt_{j-1}+\BR_{\perp}\sin\Gt_{j-1},
	\eeq{6.9}
        such that
        \beq \Bz^{\perp(j)}=\BCR_{j-1}\BR_{j-1}\Bh^{\perp(j-1)}, \eeq{6.3a}
        and because $\BR_\perp$ and $\BCR_\perp$ commute with $\BR_{j-1}$ and $\BCR_{j-1}$ we also get
        \beq \Bz^{({j-1})}=\BCR_{j-1}\BR_{j-1}\Bh^{(j-1)},\quad \Bz^{(j)}_{\perp}=\BCR_{j-1}\BR_{j-1}\Bh^{(j-1)}_\perp, \quad
        \Bz^{\perp(j)}_\perp=\BCR_{j-1}\BR_{j-1}\Bh^{\perp(j-1)}_\perp.
        \eeq{6.3b}
        In other words, $\BCR_{j-1}\BR_{j-1}$ changes the basis from $\Bh^{(j-1)}$,  $\Bh^{\perp(j-1)}$, $\Bh_{\perp}^{(j-1)}$, and $\Bh^{\perp(j-1)}_{\perp}$
        to $\Bz^{(j)}$, $\Bz^{\perp(j)}$, $\Bz_{\perp}^{(j)}$, and $\Bz^{\perp(j)}_{\perp}$ and \eq{6.4j} is satisfied with
        \beq \BQ^{({j-1})}=\BCR_{j-1}\BR_{j-1}. \eeq{6.3c}

 We again have the flexibility of replacing $\BGc$ at each stage $j$ with $\BGc_{\phi_j}$ 
given by \eq{6.4k} and \eq{6.4l} where the angles $\phi_j$ may be freely chosen. This is illustrated in Figure 3(b) and 
corresponds to allowing the orientations of the pure crystal in the sequential laminate to be uncorrelated with the directions $\Bn_j$ of lamination. 

For a general polycrystal $\Bv_\perp^{(j)}$ has a dependence on $\Bx$. For any $j$ 
one has that $\BGc_{\phi_j}\Bv_\perp^{(j)}=\Bv_\perp^{(j)}$, or equivalently
\beq \BGc_{\phi_j}(\Bx)\Bv^{(j)}_\perp(\Bx)=\Bv^{(j)}_\perp(\Bx). \eeq{6.3d}
While this  implies that $\Bv_\perp(\Bx)$ is rank 1 for each $\Bx$, there seems to be no reason 
why   $\Bz^{(j)}_{\perp}$ and $\Bh^{(j-1)}_\perp$ should both be represented by symmetric rank one matrices for all $j$ for some choice of  angles  $\phi_j$.   If this were the case then $\BQ^{(j-1)}$ would have the factorization \eq{6.3c} and a correspondence with sequential laminates could be made.

In summary, with an
                 appropriate choice of bases, solving \eq{5.11} is somewhat similar to obtaining a continued fraction expansion  for the effective tensor of a sequential laminate. The 
                 important distinction is that the change of basis requires a $\BQ^{(j-1)}$ that needs to have the form \eq{6.3c}. Whether this is possible for some choice of angles $\phi_j$ is an open problem. If it is always possible then the function $\BL_*(\BL^{(0)})$ for a general polycrystal can be mimicked, to an arbitrary high degree of approximation, by the function for a sequential laminate.

\section{Alternative continued  fraction expansions for sequential laminates}
%%%%%%%%%%%%%%%%%%%%%%%%%%%%%%%%%%%%%%%%%%%%%%%%%%%%%%%%%%%%%%%%%%%%%%%%%%%%%
\setcounter{equation}{0}
Naturally it follows from our analysis that \eq{11.1aa}, \eq{11.3aa}, \eq{6.4j}, and \eq{6.3c} provide a continued fraction expansion for the effective tensor of sequential laminates.
There are many alternative expansions based on
formulae giving the effective tensor $\BL_*$ of a laminate of two phases with tensors $\BL_1$ and $\BL_2$. For conductivity, with one phase being isotropic an elegant formula was obtained by Tartar \cite{Tartar:1985:EFC}, generalized to elasticity by Frankfort and Murat \cite{Francfort:1986:HOB}, and subsequently formulated for a wide variety of problems. For two dimensional elasticity, with one phase being isotropic a more concise formula can be obtained \cite{Gibiansky:1987:MCE}. 

One lamination formula \cite{Milton:1990:CSP, Zhikov:1991:EHM} that is convenient for developing continued fraction expansions for sequential laminates
(and underpins the theory of exact relations in composites, as reviewed in Chapter 17 of
\cite{Milton:2002:TOC}, \cite{Grabovsky:2004:AGC} and \cite{Grabovsky:2016:CMM} ) is
	\beq [\BK_*-\BGG_1(\Bn)]^{-1}=\lang[\BK-\BGG_1(\Bn)]^{-1}\rang, \eeq{17.1}
	where the angular brackets denote a volume average,
	\beq \BK_*=\Gs_0(\Gs_0\BI-\BL_*)^{-1}, \quad  \BK(\Bx)=\Gs_0(\Gs_0\BI-\BL(\Bx))^{-1}, 
	\eeq{17.2}
	and in our setting 
	\beq \BGG_1(\Bn)=\bpm \Bn\otimes\Bn & 0 \cr 0 & \Bn\otimes\Bn \epm.
	\eeq{17.2a}
	Here we are free to choose the constant $\Gs_0$ and for a given $\BL(\Bx)$, $\BL_*$ does not depend on the choice. 
	Applying this formula to a two-phase sequential laminate gives
	\beq [\BK_*^{(j)}-\BGG_1(\Bn_j)]^{-1}=f_j[\BK^{(j)}-\BGG_1(\Bn_j)]^{-1}+(1-f_j)[\BK_*^{(j-1)}-\BGG_1(\Bn_j)]^{-1},
	\eeq{17.3}
	where
	\beq 
	\BK_*^{(j)}=\Gs_0(\Gs_0\BI-\BL^{(j)}_*)^{-1},\quad
	\BK^{(j)}=\BR_j\BCR_j\Gs_0(\Gs_0\BI-\BL^{(0)})^{-1}\BCR_j^T\BR_j^T,
	\eeq{17.3a}
	in which now
	\beq \BCR_j=\BI\cos\Gy_j+\BCR_{\perp}\sin\Gy_j,\quad \BR_j=\BI\cos\Gy_j+\BR_{\perp}\sin\Gy_j.
	\eeq{17.3b}
	The directions $\Bn_j$ of lamination need not be correlated in any way with the set of crystal orientations $\Gy_j$. 
	Equivalently, we may write
	\beq \BK_*^{(j)}=\BGG_1(\Bn_j)
		+\left\{f_j[\BK^{(j)}-\BGG_1(\Bn_j)]^{-1}+(1-f_j)[\BK_*^{(j-1)}-\BGG_1(\Bn_j)]^{-1}\right\}^{-1}.
		\eeq{17.4}
	To develop the continued fraction expansion starting with $\BL_*^{(m)}=\BL_*$ one sets
	\beq \BL_*= \Gs_0\BI-\Gs_0[\BK_*^{(m)}]^{-1},
		\eeq{17.5}
	and then recursively makes the substitutions \eq{17.3}, eliminating $\BK_*^{(j)}$ for $j=m,m-1,
	\ldots, 1$ as one goes, until at the last stage one substitutes
	\beq \BK_*^{(0)} =\Gs_0\BR_0\BCR_0(\Gs_0\BI-\BL^{(0)})^{-1}\BCR_0^T\BR_0^T.
	\eeq{17.6}
	Unless there is a correspondence with sequential laminates it seems that for general polycrystals there is no such continued fraction expansion for $\BL_*$  having an analogous simple form. 
	A more complicated formula is available that simplifies to \eq{17.1} for laminates, but it does not simplify in general: see 
	\cite{Grabovsky:2000:ERE}.

	\section*{Acknowledgments}
	Mihai Putinar is thanked for his interest in the problem and for many helpful comments and for spotting a gap in the original argument. 
	The author is also grateful to the National Science Foundation for support through grant  DMS-2107926. 
	
%	\bibliographystyle{plain}
	%\bibliography{/u/ma/milton/newref, /u/ma/milton/tcbook}
%\bibliography{/Users/milton/newref, /Users/milton/tcbook}
%	\bibliography{/home/milton/newref, /home/milton/tcbook}
\section*{Appendix: Truncating the Hilbert Space}
 %%%%%%%%%%%%%%%%%%%%%%%%%%%%%%%%%%%%%%%%%%%%%%%%%%%%%%%%%%%%%%%%%%%%%%%%%%%%%
 \setcounter{equation}{0}
 Consider a domain $\CD(\Ga, \Gb, \phi)$ of tensors $\BL^{(0)}$ of the pure phase, such that for all complex
 matrices $\BA$,
 \beq \Real(e^{i\phi}\BL^{(0)}\BA, \BA) \geq \Ga|\BA|^2,\quad |\BL^{(0)}\BA|\leq \Gb|\BA|, \eeq{a.000}
 in which $\phi\in (0,2\pi)$, $\Ga$ and $\Gb$ are fixed constants, with $\Gb\geq \Ga$, $|\BA|=\sqrt{(\BA,\BA)}$, and the inner product between any pair $\BA$ and $\BB$ of complex matrices is taken to be $(\BA,\BB)=\Tr(\BA\BB^\dagger)$ where $\BB^\dagger$ is the adjoint 
 (transpose of the complex conjugate) of $\BB$. The domain  $\CD(\Ga,\Gb, \pi/2)$ 
 encompasses most tensors $\BL^{(0)}$ of physical interest which are dissipative in the sense that the self-adjoint part of  $\Imag\BL^{(0)}$ is positive definite. As mentioned in Section 4,
 the $Z$-problem will have unique solutions for each $\Be_0\in\BCU$ if $\BL_0\in\CD(\Ga,\Gb, \phi)$ 
 for some $\Ga$ and $\Gb$ and $\phi$. In order to directly apply the analysis of  section 2.4 of \cite{Milton:2016:ETC}, let us, without loss of generality, redefine $\BL^{(0)}$ to be a  suitable rotation in the complex plane of the old $\BL^{(0)}$ such that \eq{a.000} holds with $\phi=0$ implying
 the standard coercivity condition that $\Real(\BL^{(0)}\BA, \BA) \geq \Ga|\BA|^2$.

 Here we show how the infinite-dimensional Hilbert space can be truncated to a finite-dimensional one with little change
 to the effective tensor function $\BL^*(\BL^{(0)})$ in the domain $\CD(\Ga, \Gb, 0)$ of   tensors $\BL^{(0)}$ of the pure phase. 
 The proof is based upon that in Section 3 of \cite{Clark:1994:MEC}. The basic
 idea is to show that the Hilbert space can be truncated in such a way that the coefficients in the 
 series expansion of $\BL^*(\BL^{(0)})$ about the point $\BL^{(0)}=\Gs_0\BI$ remain unchanged
 up to an arbitrarily large order in the expansion for an appropriate real value of the constant $\Gs_0$.
 (If we had not taken $\phi=0$, $\Gs_0$ would need to be replaced by  $e^{i\phi}\Gs_0$.)
  The sequence of fields
 \beq \BE^{(m)}=\sum_{j=0}^m[\BGG_1(\BI-\BL/\Gs_0)]^j\Be_0,\text{   where  } \Be_0=\BGG_0\BE,
 \eeq{t.1}
 converge to the solution $\BE$ of \eq{b.4} when, for example, $\Gs_0=\Gb^2/\Ga$ and then 
 the associated sequence of tensors
 \beq \BL_*^{(m)}=\Gs_0\BGG_0+\sum_{j=0}^m\BGG_0(\BL-\Gs_0\BI)[\BGG_1(\BI-\BL/\Gs_0)]^j\BGG_0
 \eeq{t.3}
 converge to the effective tensor $\BL_*$ as $m\to\infty$.
 These convergences are proved, for example, in section 2.4 of \cite{Milton:2016:ETC}, without assuming that $\BL$ is self-adjoint. Note that
 \beqa \BE^{(m)} & = & \Be_0+[\BGG_1(\BI-\BL/\Gs_0)]\BE^{(m-1)}\text{  for } m\geq 1, \nonum
 \BE^{(0)} & = & \Be_0.
 \eeqa{t.3a}
 We now truncate our Hilbert space to a finite dimensional space by
 changing $\BGG_1$ so that $\BE^{(m)}$ remains unaltered for  $m\leq M$ as $\Be_0$ and $\BL^{(0)}$ vary.
Then the approximation given by \eq{t.3} will be close to the original effective tensor and close to the effective tensor in the modified Hilbert space. Both effective tensors become arbitrarily close to the approximation , and hence the effective tensors become arbitrarily close to each other, as  $m\to\infty$ and $M\to\infty$ while keeping $m<M$. 
 
 Let us relabel the spaces so
 \beq \underline{\BCH}=\BCU\oplus\underline{\BCE}\oplus\underline{\BCJ}
 \eeq{t.0}
 is the actual physical infinite-dimensional Hilbert space of interest, where we have introduced underlines on the spaces to distinguish them from
 the truncated spaces which we now denote as $\BCH$, $\BCE$, and $\BCJ$ with ${\BCH}=\BCU\oplus{\BCE}\oplus{\BCJ}$. We will label
 $\underline{\BGG}_1$ and $\underline{\BGG}_2$ as the projections onto $\underline{\BCE}$ and $\underline{\BCJ}$ as we will
 need slightly different operators $\BGG_1$ and $\BGG_2$ when we define the truncated Hilbert space. However,
 we still use the same notation for $\BGc$, $\BR_\perp$, and $\BCR_\perp$ as ${\BCH}$ will be closed under their action, giving the same results as these operators
 acting on the same fields in $\underline{\BCH}$. 
 
 Let us label the 16 operators that enter the expression \eq{2.15} for $\BL$:
 
\beqa \BB_1& = & \BGc,\quad \BB_2=\BGc\BR_\perp,\quad \BB_3=\BGc\BCR_\perp,\quad \BB_4=\BGc\BR_\perp\BCR_\perp, \nonum
      \BB_5& = &\BR_\perp\BGc, \quad \BB_6=\BR_\perp\BGc\BR_\perp, \quad\BB_7=\BR_\perp\BGc\BCR_\perp, \quad\BB_8=\BR_\perp\BGc\BR_\perp\BCR_\perp, \nonum
      \BB_9& = &\BCR_\perp\BGc,\quad  \BB_{10}=\BCR_\perp\BGc\BR_\perp,\quad \BB_{11}=\BCR_\perp\BGc\BCR_\perp, \quad \BB_{12}=\BCR_\perp\BGc\BR_\perp\BCR_\perp, \nonum
      \BB_{13}& = &\BCR_\perp\BR_\perp\BGc,\quad\BB_{14}=\BCR_\perp\BR_\perp\BGc\BR_\perp,\quad \BB_{15}=\BCR_\perp\BR_\perp\BGc\BCR_\perp, \quad
\BB_{16}=\BCR_\perp\BR_\perp\BGc\BR_\perp\BCR_\perp. \nonum &~&
\eeqa{t.4}
Using \eq{2.13a} the product $\BB_i\BB_j$ of any pair of these operators is either zero or equals $\pm\BB_k$ for some $k$. Also for any $i$, $\BR_\perp\BB_i=\pm\BB_k$ and
 $\BCR_\perp\BB_i=\pm\BB_\ell$ for some $k$ and $\ell$ dependent on $i$.
 
We take the four fields
\beq \BU_1=\Bt,\quad \BU_2=\BR_\perp\Bt,\quad \BU_3=\BCR_\perp\Bt,\quad \BU_4=\BCR_\perp\BR_\perp\Bt \eeq{t.5a}
as a basis for $\BCU$. Let us next introduce the multi-index fields
 \beqa \BE_{\Ga_m j} & = &\underline{\BGG}_1\BB_{a_1}\underline{\BGG}_1\BB_{a_2}\underline{\BGG}_1\BB_{a_3}\ldots\underline{\BGG}_1\BB_{a_m}\BU_j, \nonum
 \BJ_{\Ga_m j} & = & \underline{\BGG}_2\BB_{a_1}\underline{\BGG}_2\BB_{a_2}\underline{\BGG}_2\BB_{a_3}\ldots\underline{\BGG}_2\BB_{a_m}\BU_j,
 \eeqa{t.5}
 where $\Ga_m=(a_1,a_2,a_3,\ldots,a_m)$ is a multi-index comprised of indices $a_1,a_2,a_3,\ldots,a_m$ in which $m$ will be called
 the order of $\Ga_m$. Thus for a given order $m$, $\Ga_m$ can take $16^m$ different values and we denote the set of these values as $\CA_m$. 
 We define the subspaces
 
 \vskip4mm
 
 $\bullet$ $\tilde{\BCE}=$ the space spanned by the fields $\BE_{\Ga_m j}$, $j=1,2,\ldots,16$, as $a_m$ ranges over all combinations in $\CA_m$ and
 $m$ ranges from $1$ to some maximum value $m=M$. Note that $\tilde{\BCE}$ is closed under the action of $\BGY$ because $\BGY$ commutes with $\underline{\BGG}_1$ and $\BGY\BB_{a_1}$ is a linear combination of the $\BB_i$, $i=1,2,\ldots 16$. 
 
 \vskip4mm
 
 $\bullet$ $\tilde{\BCJ}=$ the space spanned by the fields $\BJ_{\Ga_m j}$, $j=1,2,\ldots,16$, as $a_m$ ranges over all combinations in $\CA_m$ and
 $m$ ranges from $1$ to some maximum value $m=M$. Similarly, $\tilde{\BCJ}$ is closed under the action of $\BGY$. 
 
 \vskip4mm
 
 The space $\tilde{\BCH}\equiv \BCU\oplus\tilde{\BCE}\oplus\tilde{\BCJ}$ is closed under the action of $\underline{\BGG}_1$, $\underline{\BGG}_2$, $\BR_\perp$ , $\BCR_\perp$ and $\BGY$ but not under the action of
 the $\BB_i$, $i=1,2,..,16$. To see this, notice that for $\BE_{\Ga_m j}\in\tilde{\BCE}$ the field
 \beqa &~& \! \!  \! \! \! \! \! \! \!\BB_i\BE_{\Ga_m j}=(\BGG_0+\underline{\BGG}_1+\underline{\BGG}_2)\BB_i\BE_{\Ga_m j} \nonum
 &~&  \quad \quad =
  \BGG_0\BB_i\BE_{\Ga_m j}+\BE_{\Gb_{m+1} j} + \nonum 
 &~&  \! \! \! \! \! \! \! \! \!
 \underline{\BGG}_2\BB_i(\BI-\BGG_0-\underline{\BGG}_2)\BB_{a_1}(\BI-\BGG_0-\underline{\BGG}_2)\BB_{a_2}
 \ldots(\BI-\BGG_0-\underline{\BGG}_2)\BB_{a_m}\BU_j, \nonum
 &~&  \quad \quad =
 \BGG_0\BB_i\BE_{\Ga_m j}+\BE_{\Gb_{m+1} j} \pm \nonum 
 &~&  \! \! \! \! \! \! \! \! \!
 \underline{\BGG}_2\BB_\ell(\BI-\BGG_0-\underline{\BGG}_2)\BB_{a_2}
 \ldots(\BI-\BGG_0-\underline{\BGG}_2)\BB_{a_m}\BU_j - \sum_{k=1}^4\Gg_k\underline{\BGG}_2\BB_{i}\BU_k -\nonum
 &~&  \! \! \! \! \! \! \! \! \!
 \underline{\BGG}_2\BB_i\underline{\BGG}_2\BB_{a_1}(\BI-\BGG_0-\underline{\BGG}_2)\BB_{a_2}
 \ldots(\BI-\BGG_0-\underline{\BGG}_2)\BB_{a_m}\BU_j, \nonum
 &~&  \quad \quad = \quad\ldots\ldots\ldots
 \eeqa{t.6}
  lies in $\tilde{\BCH}$ for $m<M$ but not generally for $m=M$, where here
  $\Gb_{m+1}$ is the multi-index $\Gb_{m+1}=(i,a_1,a_2,a_3,\ldots,a_m)$, $\BB_\ell=\pm\BB_{i}\BB_{a_1}$, and
  $\Gg_k$ is such that
  \beq \sum_{k=1}^4\Gg_k\BU_k =\BGG_0\BB_{a_1}(\BI-\BGG_0-\underline{\BGG}_2)\BB_{a_2}
  \ldots(\BI-\BGG_0-\underline{\BGG}_2)\BB_{a_m}\BU_j.
  \eeq{t.6a}
  Similarly,
 $\BB_i\BJ_{\Ga_m j}$ lies in $\tilde{\BCH}$ for $m<M$ but not generally for $m=M$. The fields in $\tilde{\BCH}$ are precisely those that appear in
 the series expansions up to order $M$ for the fields $\BE(\Bx)$ and $\BJ(\Bx)$ that solve \eq{2.11} and \eq{2.12} when $\BL^{(0)}$ is
 close to the identity matrix $\BI$, and this motivates their introduction. Note that we do not assume the set of fields $\BE_{\Ga_m j}$ (nor $\BJ_{\Ga_m j}$)
 are independent, i.e., some could be linear combinations of the others.

 Let $\BGL$ denote the projection onto $\tilde{\BCH}$, which commutes with $\BGY$,  and define $\BCW$ as the subspace spanned by the fields
 \beqa \Bw_{i\Ga_M j}& = & (\BI-\BGL)\BB_i\BE_{\Ga_M j},
 \eeqa{t.7}
 as $i$ and $j$ vary in the set  $\{1,2,\ldots,16\}$ while $\Ga_M$ varies in $\CA_M$. Note that $\BCW$ is closed under the action of
 $\BR_\perp$, $\BCR_\perp$, and $\BGY$.  Also, from \eq{2.22!}, the subspaces $\BGY\BCW$ and 
 $(\BI-\BGY)\BCW$ have equal dimension. 
 Using a Gram-Schmidt orthogonalization type process, we now find orthogonal subspaces $\BCW_\CE$ and $\BCW_\CJ$,
 each closed under the action of $\BCR_\perp$ and $\BGY$, such that
 \beq \BCW=\BCW_\CE\oplus\BCW_\CJ,
\eeq{t.7a}
and
\beq \BR_\perp\BCW_\CE=\BCW_\CJ,\quad \BR_\perp\BCW_\CJ=\BCW_\CE,\quad
\BCR_\perp\BCW_\CE=\BCW_\CE,\quad \BCR_\perp\BCW_\CJ=\BCW_\CJ.
\eeq{t.8}
Specifically, we start with one field $\Bw_0\in\BGY\BCW_0$  with $\BCW_0\equiv\BCW$ and define $\BCW_1$ as the
orthogonal complement in $\BCW$ of the subspace spanned by the four fields $\Bw_0$,  $\BCR_\perp\Bw_0$,
$\BR_\perp\Bw_0$ and $\BCR_\perp\BR_\perp\Bw_0$. We then pick a field $\Bw_1\in \BGY\BCW_1$.
 In general, we define  $\BCW_m$ as the
orthogonal complement in $\BCW$ of the subspace spanned by the $4m$ orthonormal fields $\Bw_j$,  $\BCR_\perp\Bw_j$,
$\BR_\perp\Bw_j$ and $\BCR_\perp\BR_\perp\Bw_j$ with $j=0,1,\ldots, m-1$, and we pick a field $\Bw_m\in \BGY\BCW_m$.
The process is continued until $\BCW_N$ is empty for some $N$, i.e. $\BCW$ is completely spanned by the fields that we have
generated.

We then choose $\BCW_\CE$ as the subspace spanned by the 2N fields $\Bw_j$ and  $\BCR_\perp\Bw_j$, for $j=0,1,\ldots, N-1$,
and we choose  $\BCW_\CJ$ as the subspace spanned by the 2N fields $\BR_\perp\Bw_j$ and $\BCR_\perp\BR_\perp\Bw_j$, for $j=0,1,\ldots, N-1$.
Then \eq{t.7a} and \eq{t.8} hold. Also the construction ensures that $\BCW_\CE$ and  $\BCW_\CJ$ are each closed under the action of $\BCR_\perp$ and $\BGY$.

Now we let
 \beq \BCE= \tilde{\BCE}\oplus\BCW_\CE, \quad  \BCJ= \tilde{\BCJ}\oplus\BCW_\CJ.\eeq{t.9}
 It is clear that $\BCU$, $\BCE$, and $\BCJ$ are mutually orthogonal and 
 \beq \BCH=\BCU\oplus\BCE\oplus\BCJ
 \eeq{t.11}
 defines our truncated Hilbert space. The projections $\BGG_1$ and $\BGG_2$ that project onto $\BCE$ and $\BCJ$, respectively,
 differ slightly from $\underline{\BGG}_1$ and $\underline{\BGG}_2$ (and do not act locally in Fourier space). The actions of
 $\BGc$, $\BR_\perp$, $\BCR_\perp$ on any field in $\BCH$ remain the same as the actions of these operators on the field
 before the truncation. Thus, $\BB_i$, $i=1,2,\ldots,16$, given by \eq{t.4}, and the associated operator $\BL$ acting on any field in $\BCH$
  are the same as before truncation. 
  
  The definition of $\BCE$ and $\BCJ$ ensures that the fields $\BE^{(m)}$ remain unaltered for
  $m< M$ thus ensuring that the effective tensor $\BL_*$ is almost unchanged for all 
  $\BL_0 \in \CD(\Ga,\Gb,0)$.

\end{document}